\documentclass[structabstract]{aa}
\usepackage[latin1]{inputenc}
\usepackage{txfonts}
\usepackage[pdftex]{graphicx} %
\usepackage[pdftex, plainpages=false]{hyperref}
\usepackage{natbib}
\bibpunct{(}{)}{;}{a}{}{,}

\def\kms {{\mathrm{km}\,\mathrm{s}^{-1}}}
\def\ms {{\mathrm{m}\,\mathrm{s}^{-1}}}

\def\teff {{T_{\mathrm{eff}}}}
\defcitealias{Pereira2009a}{Paper II}

\begin{document}

\title{Oxygen lines in solar granulation}
\subtitle{I. Testing 3D models against new observations with high spatial and spectral resolution}
\titlerunning{{Oxygen lines in solar granulation. I.}}

\author{T. M. D. Pereira\inst{1,2}, D. Kiselman\inst{2}, M. Asplund\inst{3}}
\authorrunning{Pereira et al.}

\institute{Research School of Astronomy and Astrophysics, Australian National University, Cotter Rd., Weston, ACT 2611, Australia \\\email{tiago@mso.anu.edu.au} 
\and The Institute for Solar Physics of the Royal Swedish Academy of Sciences, AlbaNova University Center, 106 91 Stockholm, Sweden
\and Max-Planck-Institut f\"ur Astrophysik, Postfach 1317, D--85741 Garching b. M\"unchen, Germany
}

\date{Received 5 July, 2009 / Accepted 1 September 2009}

\abstract {} {We seek to provide additional tests of the line formation of theoretical 3D solar photosphere models. In particular, we set out to test the spatially-resolved line formation at several viewing angles, from the solar disk-centre to the limb and focusing on atomic oxygen lines. The purpose of these tests is to provide additional information on whether the 3D model is suitable to derive the solar oxygen abundance. We also aim to empirically constrain the NLTE recipes for neutral hydrogen collisions, using the spatially-resolved observations of the O\,\textsc{i} 777 nm lines.} {Using the Swedish 1-m Solar Telescope we obtained high-spatial-resolution observations of five atomic oxygen lines (as well as several lines for other species, mainly Fe\,\textsc{i}) for five positions on the solar disk. These observations have a high spatial (sub-arcsecond) and spectral resolution, and a continuum intensity contrast up to 9\% at 615 nm. The theoretical line profiles were computed using the 3D model, with a full 3D NLTE treatment for oxygen and LTE for the other lines.} {At disk-centre we find an excellent agreement between predicted and observed line shifts, strengths, FWHM and asymmetries. At other viewing angles the agreement is also good, but the smaller continuum intensity contrast makes a quantitative comparison harder. We use the disk-centre observations we constrain $S_\mathrm{H}$, the scaling factor for the efficiency of collisions with neutral hydrogen. We find that $S_\mathrm{H}=1$ provides the best match to the observations, although this method is not as robust as the centre-to-limb line variations to constrain $S_\mathrm{H}$.} {Overall there is a very good agreement between predicted and observed line properties over the solar granulation. This further reinforces the view that the 3D model is realistic and a reliable tool to derive the solar oxygen abundance.}

\keywords{Sun:~granulation -- Line: formation -- Sun:~photosphere -- Techniques: spectroscopic -- Techniques: high angular resolution}

\maketitle

\section{Introduction}

One of the significant problems in modeling photospheres is how to accurately treat convection. Three-dimensional time-dependent hydrodynamical simulations of convection in solar and stellar photospheres are a major leap forward in the treatment of convection and realistic physics. They challenge the classical approximations: plane-parallel photospheres, static hydrostatic equilibrium and the mixing-length theory of convection. 

Employing the family of convection simulations of \citet{SteinNordlund1998}, the series of papers started by \citet{Asplund2000} has undertaken a systematic study of solar line formation and elemental abundances. The solar abundances derived from these models \citep{AGS05,Asplund2009} represent a downward revision of the solar metalicity by almost a factor of two. The significance of this revision echoes over many fields of astrophysics, because the Sun is a reference in abundance studies. In solar physics it has sparked much debate, mostly because the excellent agreement between the solar interior models and measurements from helioseismology is lost when using the revised solar abundances \citep{Bahcall2005}. Despite numerous tries \citep[see review of][]{Basu2008}, no work has successfully reconciled this excellent agreement with the new solar abundances.

Realistic physics, absence of free-parameters governing turbulence and the excellent observational agreement of predicted spectral line shapes \citep{Asplund2000} are some of the strong points in favour of the 3D models. Yet the disagreement with solar interior models and helioseismology has driven the question: are the 3D models suitable to abundance analysis? 

This work is part of an effort to addresses several observational tests in order to answer this question. In particular, we use high-spatial-resolution observations to test the 3D model and line formation over the solar granulation, and at different viewing angles. A robust test of the models can be established by comparing the observed and predicted properties of spectral lines over the range of temperatures and velocities found in the solar granulation. Looking at different angles (or positions on the solar disk) one can also probe higher photospheric layers and non-vertical velocity components. Such testing has been briefly discussed by \citet{Asplund2005}. The series of papers started by \citet{Asplund2000} shows that the 3D models reproduce the observed mean spatial line profiles very well. In this work we aim at resolving those mean line profiles and test the 3D models for a few lines.

We take a special interest in oxygen lines. Oxygen is pivotal in the solar abundance revision. Not only because it is abundant but also because abundances of several other elements are determined by their ratios to oxygen. After the work of \mbox{\citet{Asplund2004}} several new determinations of the solar photospheric abundance have been made: either using a different family of 3D models \citep{Caffau2008,Ayres2008} or different methods \citep{SocasNavarro2007,CentenoSocasNavarro2008}. The discussion provided by these studies prompts us to revisit the spectral indicators of oxygen. We propose to study some of the oxygen lines used in the abundance analysis. High-spatial-resolution observations of five atomic oxygen lines were obtained for this purpose. Observations were also obtained for other lines close to the target oxygen lines, and we have included some of them in our analysis (mainly Fe\,\textsc{i} lines, see Table~\ref{tab:lines}).

This work is closely related to \citet[hereafter \citetalias{Pereira2009a}]{Pereira2009a}, which focuses on the centre-to-limb variation of oxygen lines. We make use of the same observations, but here we focus on high-spatial-resolution, whereas \citetalias{Pereira2009a} studies the spatially and temporally averaged spectra as a function of viewing angle. The discussion here is primarily focused on the disk-centre granulation, because at disk-centre the contrast in intensity between granules and intergranular regions is highest. This greater contrast gives more information about the variation of spectral lines than at other viewing angles.

To study the line parameters we employ a similar methodology as \citet{Kiselman1994}. We study the variation of line strengths, shapes and shifts over the observed granulation and compare it with the 3D model. For the O\,\textsc{i} 777 nm triplet lines, known to suffer departures from LTE \citep[\emph{e.g.}][]{Eriksson1979,Asplund2005,Fabbian2009}, we seek to find whether the spatially-resolved disk-centre observations can constrain the NLTE line formation, as has previously been done using the centre-to-limb variations (\citealt{CAP2004}; \citetalias{Pereira2009a}).

The outline of this paper is as follows. In Sect.~\ref{sec:obs} we detail the observations. In Sect.~\ref{sec:red} we describe the data reduction procedure. Next, in Sect.~\ref{sec:model}, we provide details of the 3D model used, the line formation, atomic data and how the model results were degraded to the observed conditions. In Sect.~\ref{sec:results} we present the results and discuss their significance. Finally, we establish some conclusions in Sect.~\ref{sec:conc}.

\section{Observations\label{sec:obs}}
\subsection{Telescope, instruments and programme}

The observations were acquired using the Swedish 1-m Solar Telescope (SST) \citep{ScharmerSST} on Roque de Los Muchachos, La Palma. The observations took place from the 14th till the 29th of May, 2007.

Our aim was to obtain high-quality spatially-resolved spectra across the solar disk: from the disk-centre to the solar limb, with a high spatial and spectral resolution. We made use of TRIPPEL: the TRI-Port Polarimetric Echelle-Littrow spectrograph\footnote{See {\tiny http://dubshen.astro.su.se/wiki/index.php/TRIPPEL\_spectrograph}.}. The slit width is 25 $\mu$m, which translates to 0.11 arcsec. The spectral resolution ($\lambda/\Delta\lambda$) is $\approx$200\,000. TRIPPEL allows the simultaneous observation of three wavelength regions (spectrograph ports). To obtain spatially-resolved spectra along the slit, there is no predisperser. All spectral orders are overlapped and order selection is done by using filters.

The spectrograph cameras used were two MegaPlus 1.6 cameras (615 and 777 nm windows) and one MegaPlus II es1603 camera (630 nm window). In addition to the spectrograph and its three cameras, we operated four imaging cameras: three slit-jaw image cameras (at the wavelengths of 630, 694 and 773 nm, synchronized with the spectrograph's cameras) and one `blue' camera with a Ca\,\textsc{ii}~H filter, using a dichroic beam splitter for $\lambda\lesssim 500\:\mathrm{nm}$. The slit-jaw cameras allowed for structure identification; the Ca~H camera was used as a monitor of magnetic activity. The exposure times used were in the range of 30 - 265 ms, depending on the filters and the position on the solar disk. With neutral density filters to ensure identical exposure times, the slit-jaw cameras were synchronized with the spectrograph's cameras.

Our main target being oxygen lines, the filters for the three wavelength regions of the spectrograph were the following: 615.40, 630.08 and 777.52 nm (filter transmission windows are between 5.1-6.8 nm, centered on the filter wavelength). Hereafter referred as 615, 630 and 777 regions, they correspond respectively to the spectrograph orders 37, 36 and 29. These wavelength regions allowed us to observe the O\,\textsc{i} 615.81 nm line, the [O\,\textsc{i}] 630.03 nm line and the three O\,\textsc{i} triplet lines around \mbox{777 nm}. A summary of the lines of studied in these three regions is given in Table~\ref{tab:lines}. 

High-spatial and spectral resolution observations of these lines were obtained for several positions on the solar disk, from the centre to the limb. The positions in the disk are defined by $\mu\equiv \cos\theta$, where $\theta$ is the heliocentric viewing angle. Because the 3D models used do not include magnetic fields, our aim in this work is to study the quiet Sun. While observing, efforts have been made to avoid regions with obvious magnetic fields (\emph{e.g.} by using the Ca~H camera). However, magnetic fields in the Sun can never be completely avoided.

In total more than 890\,000 images were taken. The data were taken in a variety of seeing conditions, hence the same quality was not possible for all observing runs. Extensive use was made of the SST's Adaptive Optics system \citep{ScharmerSST_AO}. Some difficulties of acquiring AO lock were experienced close to the limb where the contrast of solar features is usually low. In some cases (notably $\mu = 0.4$) targets with higher contrast, like facular points, were chosen to facilitate locking. These structures were presumably magnetic features and the procedure thus somewhat compromised the effort to observe as quiet sun as possible. Another systematic difference between datasets results from the fact that the disk-centre observations had higher priority. Thus more time was spent on them and they happened to catch the moments of best seeing. 

\begin{figure} 
\begin{center}
  \includegraphics[width=0.4\textwidth]{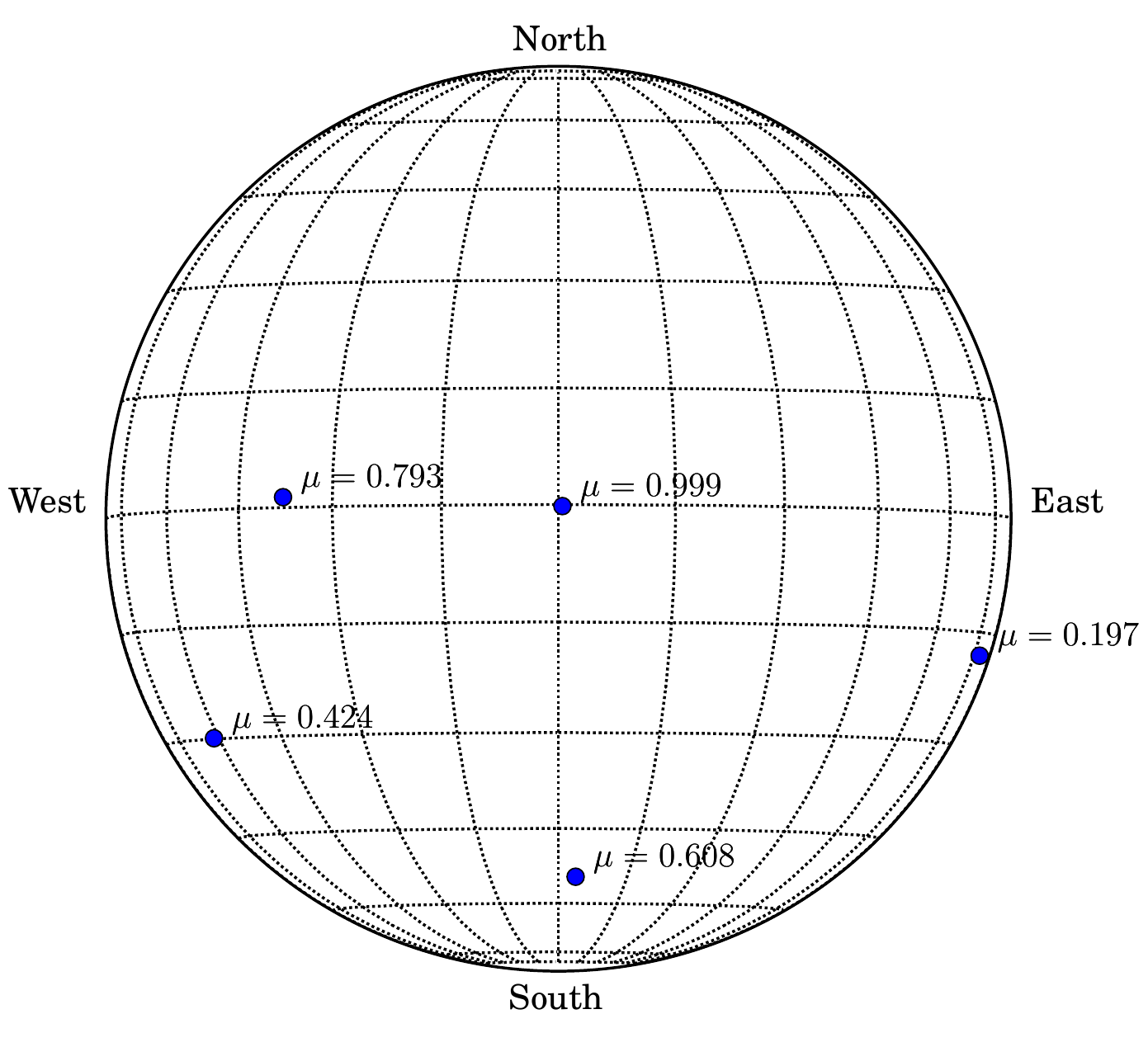}
\caption{Position of the observations on the solar disk, in heliographic Stonyhurst coordinates. The figure has been centered in the heliographic latitude of the observer ($\approx -1.8\degr$ for our dates).}
  \label{fig:obs_map}
\end{center}
\end{figure}

Details about the data and the coordinates in the Stonyhurst system \citep{Cortie1897} can be found in Table~\ref{tab:subsets}, and the observed locations on the solar disk are shown in Fig.~\ref{fig:obs_map}. For the 777~nm camera at $\mu\approx 0.8$ the dataset from the 25th of May was used. This was due to problems with a misaligned 777 nm filter for the 27th May set (the adopted set for the other cameras).

\begin{table}[ht]
\caption{Observing runs selected for analysis. Coordinates are given in the Heliographic Stonyhurst system. The spread in $\mu$ is given by the approximate spatial coverage of the slit. The $\Delta I_{\rm{rms}}$ is given as the maximum continuum contrast of a single frame in the 615 nm wavelength region.}
\label{tab:subsets}
\begin{center}
\begin{tabular}{cclc}
\hline\hline
UT Date & $\mu$ & Coordinates &  $\Delta I_{\rm{rms}}$ [\%] \\
\hline
20/5/2007 08:00 & $0.999\pm0.001$ & 0.2S, 0.5E   & 9.2 \\
25/5/2007 08:30 & $0.816\pm0.010$ & 26.7N, 22.1E & 5.4 \\
27/5/2007 15:30 & $0.793\pm0.012$ & 1.3N, 37.5W  & 6.0 \\
25/5/2007 09:56 & $0.608\pm0.020$ & 51.1S, 3.7E  & 4.7 \\
20/5/2007 09:27 & $0.424\pm0.024$ & 29.9S, 61.4W & 5.2 \\
25/5/2007 07:40 & $0.197\pm0.003$ & 18.0S, 82.4E & 3.4 \\
\hline
\end{tabular}
\end{center}
\end{table}

\subsection{Pointing accuracy}

To minimise the spread in $\mu$ for each data set, a position on
the disk was chosen so that the slit was parallel to the closest part
of the limb. The field of view is rotating in the laboratory frame
and thus the position angle of the selected locations changed during the
day. This explains the scattered appearance of the points in Fig.~\ref{fig:obs_map}.

The SST's blind pointing is not precise enough for $\mu$ to be
determined with the required precision close to the limb.
We took advantage of the Dutch Open Telescope (DOT) to help with the
coordinates. With
the kind assistance of P. S{\"u}tterlin, the DOT was pointed at the
same solar region as the SST by comparing the Ca\,{\sc ii}~H filtergram
view of the two telescopes. The heliographic coordinates
were thus read from the DOT. The spread in $\mu$ in Table~\ref{tab:subsets}
is due to the range of $\mu$ covered by the slit.

Note that the outermost images around $\mu = 0.2$ have a very
well-defined $\mu$ because the limb was in the field of view of the
slit-jaw images, which allowed precise measurements. This explains the
small spread in $\mu$ for that position.

\section{Data reduction\label{sec:red}}

\subsection{Dark currents, flat fields and spectrograph distortions}

The first step in the reduction was the subtraction of the dark currents and division by a normalized flat field image. Darks were obtained by blocking the sunlight. Because of the features on the solar surface and the presence of the spectral lines, the task of obtaining a uniformly exposed flat field needs a more elaborate approach. The telescope is operated in a special mode where it scans randomly quiet areas at the disk-centre. A reference disk-centre spectrogram is obtained by averaging these images. The spectral lines are removed by dividing this spatially uniform spectrogram by its mean spectrum (spatial average), and a flat field image is obtained\footnote{For the spectrograph's cameras. For the imaging cameras the flat-field is simply the normalized averaged of the many disk-centre images.}. This disk-centre mean spectrum is also used for wavelength, intensity and scattered light calibrations, as detailed below.

Once corrected for dark currents and flat fielded, the images need to be corrected for the spectrograph's distortions: smile and keystone. The aim is to rectify the spectrogram, to have the dispersion exactly perpendicular to the pixel rows. Smile curves the spectral lines across the spatial axis, whereas keystone curves them across the spectral axis. The distortion is computed by fitting a fourth order polynomial to the line centres of a few lines along the slit (in a spatially uniform image). This information is used to compute the transformation to invert the distortion and interpolate the spectrogram into a rectified version. The spectrogram is also slightly cropped to remove dark borders.

\subsection{Wavelength and intensity calibrations}

Wavelength calibration is made by comparison with the reference FTS atlas of \citet{BraultNeckelFTS}. Using the disk-centre mean spectrum obtained from the flat fielding procedure, for each wavelength region a few lines are identified in the mean spectrum and in the FTS atlas. Their line centres are computed and a polynomial is fitted to the dispersion relation (wavelength/pixel number) -- yielding a wavelength scale for our images. The precision of this wavelength calibration has proven to be more than enough for our purposes. When computing the line shifts across the solar granulation, velocity measurements are made relative to the reference mean spectrum (the spatial and temporal mean of the images at a given $\mu$) -- thus a very precise absolute wavelength calibration is not required. Using the telluric lines in the 630 nm region to calibrate the wavelengths more robustly we find a very good agreement ($< 40\:\ms$) with our `normal' procedure.

All spectral intensity measurements are made relative to the observed disk-centre intensity. In the flat fielding method described above, a spatially and temporally averaged reference disk-centre spectrum (hereafter referred as `mean spectrum') is obtained. The intensity of each spectrogram observed at a given position on the solar disk is then divided by the (continuum) intensity of the mean spectrum. 

For the same exposure times, the exposure level varies with time-dependent factors, most importantly airmass. Because the flat fields are also used to calibrate the intensity, it is important to have the exposure levels of flat fields and observations as close as possible. To minimize for the time-dependent factors the observations and the mean spectrum are obtained in close intervals (20-30 min). These intervals are usually enough to keep the same exposure level between observations and flat-fields (for the mean spectrum). However, in the early morning and late afternoon when the airmass varies quickly with time it is necessary to correct for the different airmasses. Our approach is empirical: using several thousand slit-jaw images for the given airmass interval, we fit the terrestrial atmospheric absorption coefficient $\tau_\oplus$ from the logarithmic intensity \emph{vs.} airmass relation. 

We assume the terrestrial atmospheric extinction to be given by:
\begin{equation}
  \label{eq:int}
  I = I_0\exp(-a\tau_\oplus),
\end{equation}
where $a$ is the airmass. The airmass correction factor is then the ratio of flat fields/observations intensity:
\begin{equation}
  \label{eq:tau}
  I_{\rm{factor}} = \exp(-\tau_\oplus [a_{\mathrm{ff}} - a_i]),
\end{equation}
where $a_{\mathrm{ff}}$ is the mean airmass when the flat fields were taken and $a_i$ the airmass when each image was taken.
The intensities from slit-jaw images are then used to derive $\tau_\oplus$ empirically. With a linear fit to the $\log(I)$ vs. airmass relation we obtain $\tau_\oplus$ and $\log(I_0)$. Due to the absence of a dedicated slit-jaw camera, for the 615 nm region $\tau_\oplus$ was extrapolated in wavelength from the other regions.

In some circumstances the camera shutters introduce a slightly different exposure between odd and even frames. This effect is corrected by scaling the exposure level of each odd/even frame by the ratio of the mean intensity to odd/even intensity.

\subsection{Correction for straylight in the spectrograph}

Light scattered off the grating of the TRIPPEL spectrograph as well as from other internal sources introduces straylight in the spectra. In our treatment for straylight, we have assumed that it is constant along the slit and is proportional to the mean intensity of light entering the spectrograph. Our treatment was based on routines used in \citet{CAP2004}, that were kindly provided by C. Allende Prieto. The relative level of straylight, given as a fraction of the continuum intensity, is computed using the FTS atlas and the calibration disk-centre mean spectrum. First the mean spectrum is normalized. Assuming that the spectrograph's instrumental profile is Gaussian, we derive the straylight level and the resolution relative to the FTS atlas by comparing our mean spectrum with the atlas (convolved with a Gaussian for an arbitrary resolution). The approach searches for the best fitting values of the straylight and the FWHM of the convolving Gaussian. Using an iterative $\chi^2$ minimization, the best fitting pair is found. This straylight and resolution represent the parameters that best `map' the FTS atlas into our mean spectrum. This procedure is repeated for each wavelength region and each data subset. We find a straylight level of $\approx$6\%, which is in agreement with other observations with this instrument \citep[\emph{e.g.}][]{Langangen2007}. This relative straylight level is then subtracted from each spectrogram.

\subsection{Normalization}

\begin{figure} 
  \centering
  \includegraphics[width=0.5\textwidth]{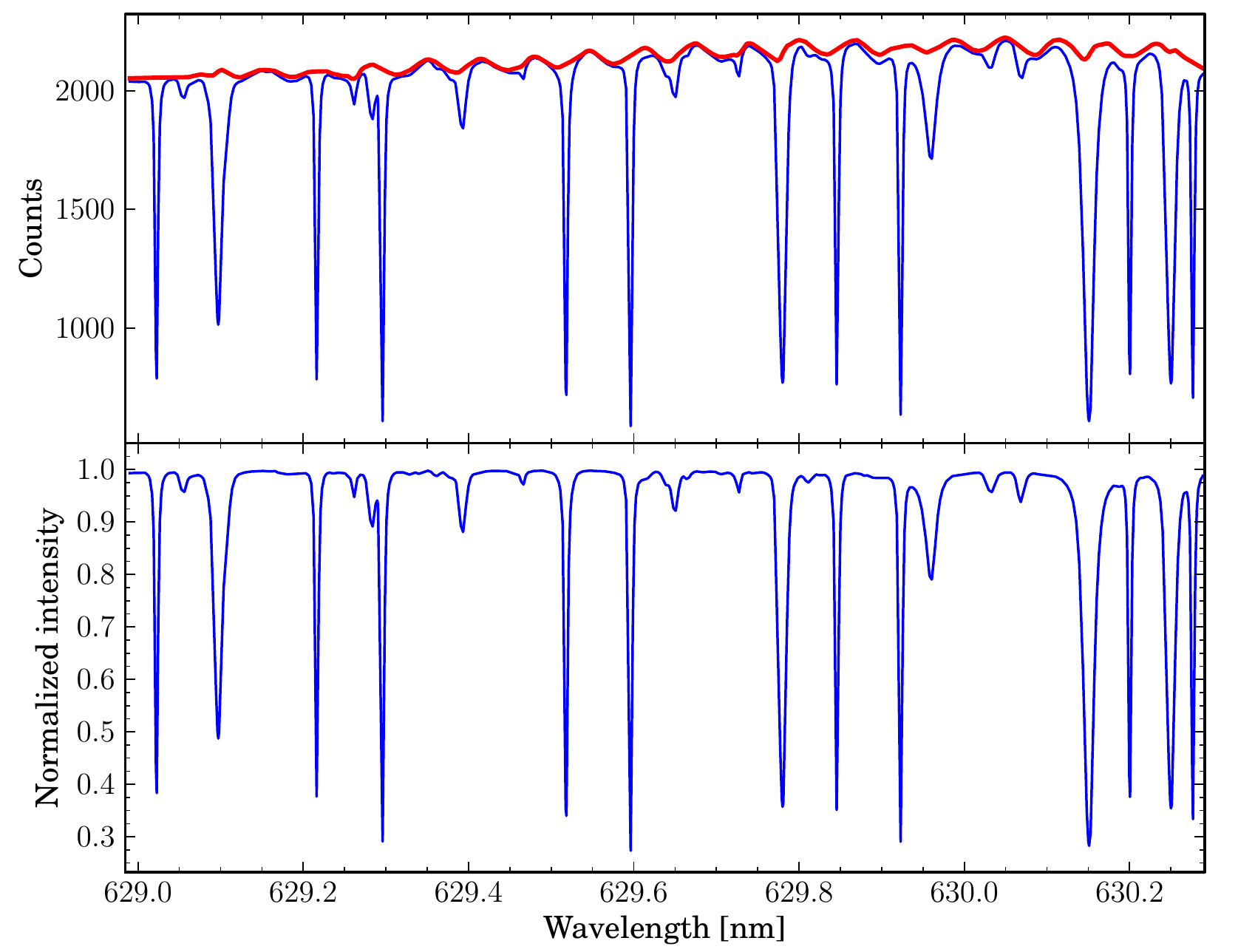}
  \caption{Normalization of the mean spectrum, for a frame in the 630 nm region affected by fringing. Top: mean spectrum before normalization (thin line), continuum map from our algorithm (thick line). Bottom: mean spectrum after normalization.}
  \label{fig:contmap}
\end{figure}

Normalization of the spectrogram is one of the most important steps of the reduction. There are several factors that determine the overall shape of the spectra: filter transmission function, grating blaze function, vignetting, fringing, etc.. Flat fielding removes most of the fringing, however our procedure will leave traces of fringe components parallel to the spectral lines. Our approach to normalization is a `one-fits-all', in which we correct for all these effects in one step. Once again, the FTS atlas is used. It is assumed that it perfectly reproduces the mean spectrum at disk-centre. Our procedure is to find the transformation that maps the continuum of our mean spectrum into the continuum of the FTS atlas. This is done by first convolving the atlas with a Gaussian instrumental profile. Our mean spectrum is then divided by the atlas and a `continuum map' is obtained by smoothing this ratio. We find that this procedure is extremely efficient at removing fringing and other artifacts. Due to the filter position the wavelength region around 630 nm happened to be the most affected by fringing. Figure~\ref{fig:contmap} gives an example of how the continuum finding procedure works in this region.

Because there is no absolute intensity calibration, normalization of the calibration mean spectrum defines the disk-centre intensity. All spectrograms are divided by the corresponding continuum map, which traces the continuum level in the mean spectrum. This ensures that all the distortions in the spectra are corrected. The next step is to find the continuum level, and for each spatial point, as it varies along the granulation pattern. This is done by identifying the continuum regions in the spectrum (defined as points where $I \gtrsim 0.995$ in the FTS atlas). The continuum level is taken as the mean intensity in such regions.

\subsection{Frame selection and sampling}

\begin{figure} 
  \centering
  \includegraphics[width=0.4\textwidth]{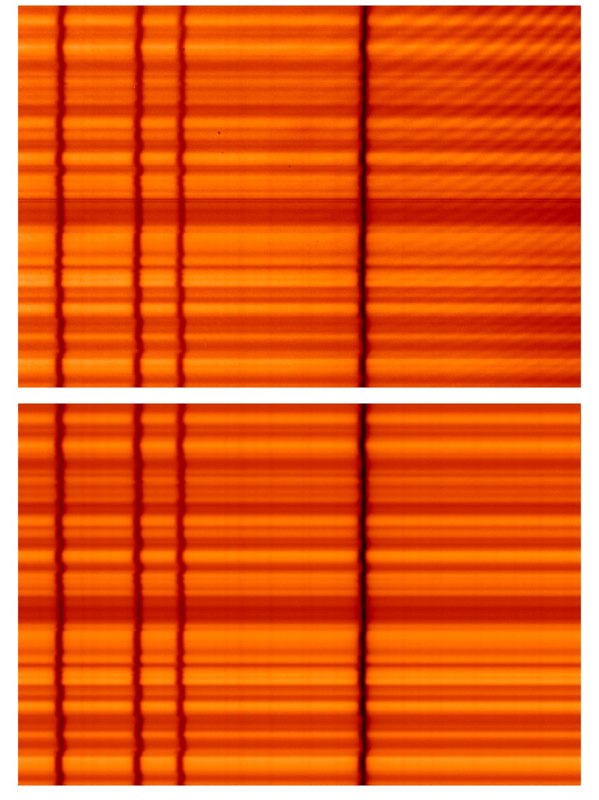}
  \caption{Example spectrogram for the 777 nm region before (top) and after (bottom) the reduction. In the final reduced image distortions and inhomogeneities such as the fringing pattern in the top right; vignetting; continuum function and dust specs in the CCD have been corrected. The vertical axis represents spatial position and the horizontal axis wavelength. The three vertical lines on the left are the O\,\textsc{i} 777 nm lines, and the darker line on the right is the strong Fe\,\textsc{i} 778.05 nm line.}
  \label{fig:raw_vs_red}
\end{figure}

In our observations more than 392\,000 spectrograms were obtained. From these, only about 500 were actually used in the analysis. The most important criterion for data selection was the continuum intensity contrast. The maximum values for the continuum contrast for the 615 nm are shown in Table~\ref{tab:subsets}. The disk-centre value of 9.2\% demonstrates the excellent quality of the data. It can be compared with other values obtained with slit spectrographs, like the $\approx 7\%$ at 615 nm of \citet{Kiselman1994} with the SVST and the $7\%$ at 630 nm\footnote{The continuum contrast is inversely proportional to wavelength (approximately). Our value of 9.2\% at 615 nm corresponds to $\approx$9.0\% at 630 nm.} obtained by \citet{Danilovic2008} with the SOT/SP of the Hinode space telescope. 

Continuum contrast is a good measure of quality for quiet Sun regions, but cannot be used blindly. Some magnetic features are difficult to identify just by visual inspection of the slit-jaw images, and can artificially increase the continuum contrast. Because the present work is concerned with quiet regions, some subsets had to be removed from the analysis because they were suspect of having magnetic effects (\emph{e.g.}, excessively bright spatial points). Other sets displayed an excessive amount of fringing that could not be completely corrected, and therefore were discarded. The best subset for each $\mu$ value was used. From each subset selected for analysis the best 25 images were used, except for disk-centre where the best 50 images were used. 

In Figure~\ref{fig:raw_vs_red} an example spectrogram is displayed, showing the effects of reduction in the image.

An important aspect of the selected frames is the temporal and spatial sampling. With the simulations we have a very good time and spatial sampling of the solar granulation. However this is not always possible with the observations if we want to select the best continuum contrast. The observations with higher contrast will tend to be clustered in the instants with better seeing, which do not provide the best sampling. There is a trade-off between getting the highest quality images and getting the best sampling. In this work we have seeked the higher quality images, at the expense of sampling.

A spatial sampling as good as the simulations will always be difficult to achieve because the models have a two dimensional sampling of the solar surface, while the spectrograph's slit is only one dimensional. Our slit covers $\approx 40\,\rm{arcsec}$, while the models cover a box of $\approx 8\times 8\,\rm{arcsec}^{2}$. In the time domain the simulations have a better coverage (20 snapshots uniformly sampled from 45 minutes of solar time). Our observations at disk-centre cover slightly more than 10 minutes, while at other viewing angles the time coverage is typically of a few minutes. The solar 5-minute oscillations will influence the line properties (mainly velocities), so a good temporal sampling is important to average the oscillations, which are also present in the simulation. For our disk-centre data the coverage is adequate for this purpose, but for other viewing angles it is not so good. The problem of systematics in the line shifts for $\mu\neq 1$ due to solar oscillations has been minimized by measuring the line shifts relative to the spatially and temporally averaged spectrum for the given $\mu$, and not from the mean disk-centre spectrum.

\subsection{Fourier filtering}

\begin{figure}
\begin{center}
  \includegraphics[width=0.5\textwidth]{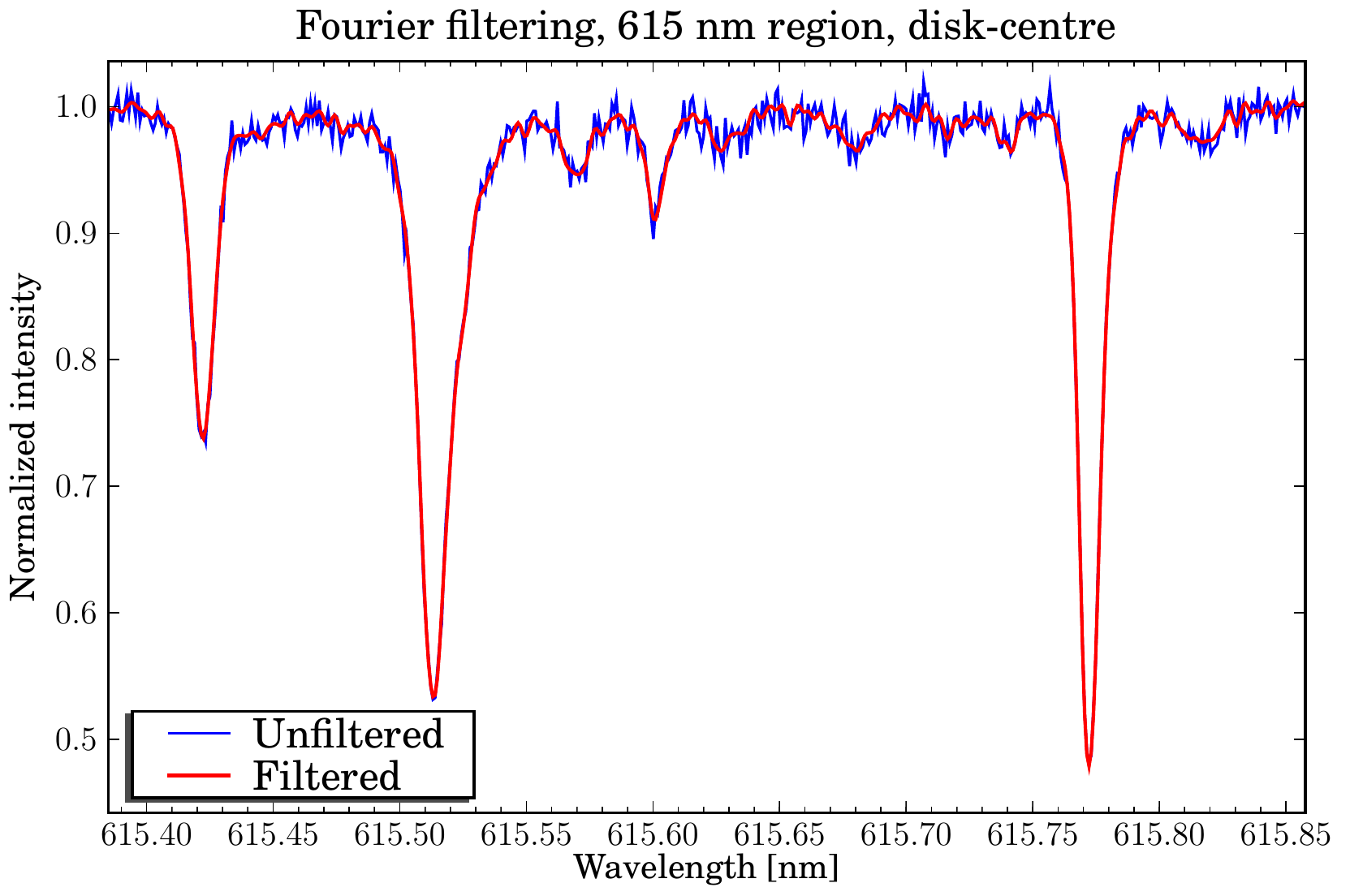}
\end{center}
\caption{Effect of the Fourier filter in a spectral region around 615 nm. The high-frequency noise components are eliminated in the filtered spectrum.}
  \label{fig:fourierfilter}
\end{figure}

The signal-to-noise ratio of each individual spatially-resolved spectrum is on the order of 25-50, depending on the camera used and the exposure level. To minimize this noise a Fourier filter was implemented. Because the spectrograms are oversampled\footnote{There are at least 5 pixels per resolution element in the spatial axis and 3 in the wavelength axis.} in the spatial and wavelength directions a Fourier filter is very effective at minimizing the noise with little loss of signal. Given that on the spectrograms noise has a much higher spatial frequency than the spectral lines, a low-pass filter eliminates the noise while preserving the spectral information. Working on the Fourier space the high frequencies are filtered out by multiplying the two dimensional Fourier transform of the spectrogram by a low-pass filter function, and taking the inverse Fourier transform to obtain the filtered data. For the filter function we use a two dimensional Butterworth filter with an elliptic shape, defined as:
\begin{eqnarray}
  \label{eq:ffilter}
D &=& \left(\frac{f_{x}}{c_x}\right)^2 + \left(\frac{f_{y}}{c_y}\right)^2 \\
F_{a,b,p} &=& \frac{1}{1+D^n},
\end{eqnarray}
where $n$ is the order of the filter, $D$ is the distance in the two dimensional frequency space ($f_{x}$, $f_{y}$), $c_x$ and $c_y$ are the cutoff frequencies in $x$ and $y$. An order of $p=20$ was used, and the chosen cutoff frequencies were $c_x=c_y=0.1\:\mathrm{pix}^{-1}$. These values ensured that a significant part of the noise was eliminated while retaining the spectral information. The smooth dip in the function ensures that Fourier artifacts such as the Gibbs phenomenon are minimized. Being a two dimensional problem, noise in both axis of the spectrogram (wavelength and spatial coordinate) can be treated differently by setting different values for $c_x$ and $c_y$.

Even with the conservatively chosen cutoff frequencies, Fourier filtering significantly increases the signal-to-noise ratio. Its effects are very helpful for a precise continuum determination and for very weak lines. Fig.~\ref{fig:fourierfilter} shows an example of the effects of the filtering.

\subsection{Extracting line quantities}

Five line properties are extracted from the spectra: line-centre position, line-centre intensity, full width at half maximum (FWHM), a measure of line asymmetry, equivalent width. These are computed automatically (i.e., no human input in the fitting procedures) using algorithms designed for the effect.

We define `line centre' as the velocity difference in $\kms$ between the line centre of the mean spectrum (spatial and temporal average the spectra at the given $\mu$ value) and the line centre for the spectrum of a given spatial point. The line centre wavelength and intensity are obtained by fitting a parabola to a few points around the line minimum. The velocity shift is then computed using the mean spectrum and converting wavelength to velocity units. The line centre intensity is given in relative intensity units.

The line FWHM is computed by fitting the line wings and measuring the width at half maximum. Two 5th order polynomials are fitted to the red and blue wings of the line being analyzed. The fitted functions are then evaluated at the half maximum of the line, and the wavelength distance between red and blue wings is measured. This distance corresponds to the FWHM, and is converted to velocity units. The quantity `line asymmetry' is computed in a similar way. The line asymmetry is defined as difference (in velocity units) between the line centre and the bisector at half maximum. The wavelength position of the line bisector at half maximum is obtained as the FWHM: from the fitted polynomials to the line wings. It is then subtracted from the line centre wavelength and converted to velocity units. Due to the fitting procedure, the FWHM and `line asymmetry' are the quantities with the highest uncertainties.

The equivalent widths were computed by direct integration. For consistency both the observations and the (convolved) simulations were run through the same analysis procedure. 

The uncertainties for all the line quantities reflect the photon noise and continuum placement errors. The effects of the photon noise, were estimated using artificial lines. To generate an artificial line with similar properties, a Gaussian was fitted to the mean spectrum of each line. This artificial line was then placed in a continuum region in a few spectrograms. Its quantities were computed in the usual fashion and the comparison between the Gaussian exact quantities with the extracted quantities yields the errors due to photon noise. The final errors were the quadratic sum of the photon noise and continuum placement errors. The continuum placement error was assumed to be 0.25\% for all lines, for each spatially-resolved spectrum.

\section{Simulations\label{sec:model}}

\subsection{Atmospheric model}
The 3D line formation calculations presented herein have been
performed with a 3D, time-dependent, radiative-hydrodynamical
simulation of the solar photosphere (Trampedach et al., in preparation;
Asplund et al., in preparation). This new 3D model differs in a number of
ways from the 3D solar atmosphere of \citet{Asplund2000} that has
been extensively used for solar abundance purposes over the past decade
\citep[\emph{e.g.}][]{CAP2004,Asplund2004,Asplund2005b}.
It has been computed with the {\sc stagger} MHD code 
\citep{NordlundGalsgaard1995,GudiksenNordlund2005}, although no
magnetic field has been imposed for the present simulation.
The code solves the hydrodynamical equations of mass, momentum and energy
coupled with a solution of the 3D radiative transfer equation using nine directions
(vertical direction plus 2 inclined with 4 azimuthal rays).
The code makes use of periodic horizontal and open vertical boundaries.
The Eulerian mesh consists of $240^3$ cells, which corresponds to a physical
dimension of 6$\times$6$\times$4\,Mm; the solar simulation extends about 900\,km above the
optical depth unity. For the subsequent line formation calculations the original
simulation was interpolated to a finer vertical depth scale while sampling
the horizontal directions to produce a 3D model of dimensions 50$\times$50$\times$82;
test calculations confirmed that this procedure did not introduce any differences
in the final results.

Compared to our previous 3D solar model, an extensive overhaul of the
input micro-physics and radiative transfer treatment has been carried out.
The continuum and line opacities are based on the {\sc marcs} code \citep{MARCS2008}
supplemented by additional photo-ionization cross-sections from the
Opacity and Iron Projects \citep{Cunto1993,Hummer1993}.
The adopted equation-of-state is that of \citet{MHDI}.
The \citet{AGS05} solar chemical composition has been used both
for the opacities and equation-of-state.
The wavelength-dependence of the opacities have been accounted
for using the concept of opacity binning \citep{Nordlund1982}. The division in
opacity and wavelength for the 12 opacity bins was based on
the full monochromatic ($105$ wavelength points) solution for the mean
3D stratification. Continuum scattering was included as a pure absorption
(i.e. $S_\lambda = B_\lambda(T)$, which is fully justified for the solar photosphere
(Hayek et al., in preparation).

The time-sequence of the 3D solar simulation utilized for the line formation calculations
corresponds to 1.2\,hr of solar time. The resulting effective temperature of the
 90 snapshots is $T_{\rm eff} = 5778 \pm 14$\,K.

\subsection{Line formation and departures from LTE\label{sec:line_formation}}

Using the 3D model described above we computed the synthetic line profiles using our LTE line formation code. From the full simulation a series of 20 snapshots covering $\approx$40 min of solar time has been used. The adopted snapshots were chosen so that the temporally averaged effective temperature of the atmosphere (not a free parameter in the simulation) was $\teff=5777\pm 14\:\mathrm{K}$, very close to the observed nominal value of $5777\pm 3\:\mathrm{K}$ \citep{WillsonHudson1988}. While the full simulation has a numerical resolution of $150\times150\times82$, for the purposes of spectral line formation we have used an interpolated lower resolution of $50\times50\times82$. As outlined in \citet{Asplund2000} this procedure is meant to save computing time and does not introduce any differences in the line profile shapes. Assuming LTE, the line formation code used a Planckian source function ($S_\nu=B_\nu$), and the Saha and Boltzmann equations to determine respectively the ionization fraction and atomic level populations for each species. For realistic equation-of-state and continuum opacities the  Uppsala opacity package was employed. The radiative transfer equation has been solved for several directions: for each of the five $\mu$ positions of our observations the line profiles were computed over for four $\varphi$ angles. For each line the profiles were computed using three abundance values, allowing for interpolation of other abundance values. No additional line broadening through micro or macroturbulence was applied.%

The synthetic line profiles were computed assuming the LTE approximation. For oxygen lines it was possible to perform 3D NLTE line formation calculations. This was accomplished using the \textsc{multi3d} code \citep{Botnen1997,Botnen1999,Asplund2003a}, which solves the rate equations iteratively using a local lambda operator and assuming statistical equilibrium. For the purposes of calculating the level populations, the radiative transfer in \textsc{multi3d} was solved for 40 outgoing inclined rays using a short characteristic technique. Once the level populations had converged and the iteration is finished, we made use of the resulting source function and opacities to compute the radiative transfer using a long characteristic Feautrier solver. Similarly to the LTE code, we compute the NLTE line profiles for the $\mu$ values of our observations, with 4 $\varphi$ angles each. The Uppsala opacity package was used to compute the continuum opacities and equation-of-state for \textsc{multi3d}. For increased numerical accuracy, the 3D model fed into \textsc{multi3d} had a resolution of 100 points in the vertical scale, interpolated from the 82 points of the original 3D simulation. 

Due to the computational expense, full 3D NLTE line formation was carried only for 10 snapshots of the simulation. The snapshots were selected to provide a good statistical coverage, both in time (evolution of granulation) and effective temperature. The final NLTE line profiles were obtained by multiplying the LTE line profiles (computed with our LTE code) by the NLTE/LTE ratio (computed with \textsc{multi3d}). The wavelength-dependent NLTE/LTE ratio was obtained for each snapshot, viewing angle and spatial point. Multiplication of the ratio by the line profiles is made individually for all spatially-resolved spectra and for each viewing angle and snapshot.

Because one has NLTE profiles directly from \textsc{multi3d}, it may seem counterproductive to obtain the final values by multiplying the LTE profiles by the NLTE/LTE ratio. However, this approach was followed for two reasons. First, to allow interpolation of the line profiles in abundance. Line profiles from \textsc{multi3d} are computed only for one abundance value, whereas for our LTE code they are computed for three values of $\log(\epsilon_{\rm{O}}\cdot gf)$. Interpolating between these three line profiles allowed us to calculate the line profile for an arbitrary oxygen abundance. The second reason this approach was taken was to ensure consistency between different lines by using the same line formation code. In any case, for the oxygen lines studied we find very little difference in the LTE line profiles between \textsc{multi3d} and our LTE code.

\subsection{Atomic data\label{sec:atomd}}

The two main sources of atomic data for our spectral calculations were the NIST database \citep{NIST} and the VALD database \citep{VALD1,VALD2}. A summary of the lines properties and the sources used can be found in Table~\ref{tab:lines}.
Collisional (van der Waals) broadening was computed using the quantum mechanical theory of \citet{AnsteeOMara1995,Barklem1997,Barklem1998}, eliminating the need for conventional \citet{Unsold1955} enhancement factors. The collisional broadening coefficients were computed for the oxygen lines and, when available, to other lines used in our analysis. 

\begin{table}[ht]
\caption{Lines studied in the present work and their parameters. Equivalent widths at disk-centre measured from our observations. Data sources: a) VALD database; b) NIST database; c) \citet{Nave1994}; d) \citet{Storey2000}.}
\label{tab:lines}
\begin{center}
\begin{tabular}{ccrrr}
\hline\hline
Atomic Species & $\lambda$ [nm] & $\log gf$ & $E_{\mathrm{low}}$ [eV] & $W_{\mu=1}$ [pm] \\
\hline
Fe\,\textsc{i} & 615.1618$^{\rm{c}}$ & -3.299$^{\rm{b}}$ &  2.18$^{\rm{b}}$ & 4.95\\
Si\,\textsc{i} & 615.5693$^{\rm{a}}$ & -2.252$^{\rm{a}}$ &  5.62$^{\rm{a}}$ & 0.61\\
Ca\,\textsc{i} & 615.6023$^{\rm{a}}$ & -2.497$^{\rm{a}}$ &  2.52$^{\rm{a}}$ & 0.92\\
Fe\,\textsc{i} & 615.7728$^{\rm{c}}$ & -1.260$^{\rm{a}}$ &  4.08$^{\rm{b}}$ & 6.45\\
O\,\textsc{i}  & 615.8186$^{\rm{b}}$ & -1.841$^{\rm{b}}$ & 10.74$^{\rm{b}}$ & 0.63\\
Fe\,\textsc{i} & 615.9378$^{\rm{c}}$ & -1.970$^{\rm{a}}$ &  4.61$^{\rm{a}}$ & 1.21\\
\hline
Fe\,\textsc{i} & 629.0965$^{\rm{a}}$ & -0.774$^{\rm{a}}$ & 4.73$^{\rm{a}}$ & 7.06\\
Fe\,\textsc{i} & 629.7792$^{\rm{b}}$ & -2.740$^{\rm{b}}$ & 2.22$^{\rm{b}}$ & 7.54\\
$$ [O\,\textsc{i}] & 630.0304$^{\rm{b}}$ & -9.717$^{\rm{d}}$ & 0.00$^{\rm{b}}$ & 0.45\\
Sc\,\textsc{ii} & 630.0698$^{\rm{a}}$ & -1.887$^{\rm{a}}$ & 1.51$^{\rm{a}}$    & 0.59\\
\hline
O\,\textsc{i}  & 777.1944$^{\rm{b}}$ & 0.369$^{\rm{b}}$ & 9.15$^{\rm{b}}$ & 8.48\\
O\,\textsc{i}  & 777.4166$^{\rm{b}}$ & 0.223$^{\rm{b}}$ & 9.15$^{\rm{b}}$ & 7.44\\
O\,\textsc{i}  & 777.5390$^{\rm{b}}$ & 0.002$^{\rm{b}}$ & 9.15$^{\rm{b}}$ & 5.90\\
Fe\,\textsc{i} & 778.0557$^{\rm{a}}$ & 0.029$^{\rm{a}}$ & 4.47$^{\rm{a}}$  & 14.81\\
\hline\hline
\end{tabular}
\end{center}
\end{table}

Aside from the lines listed in Table~\ref{tab:lines} we have included nearby blends in some line profiles. For the [O\,\textsc{i}] 630.03 nm we included the Ni\,\textsc{i} blend and for O\,\textsc{i} 615.81 nm we have included some weak molecular and atomic blends \citepalias[for more details about the blends in these lines, see][]{Pereira2009a}. In the line profiles of Si\,\textsc{i} 615.57 nm and Ca\,\textsc{i} 616.60 nm we have included both lines and also a nearby strong Si\,\textsc{i} line at 615.51 nm. In the case of Ca\,\textsc{i} 616.60 nm we have also included the very weak O\,\textsc{i} triplet at 615.60 nm. Data for all these lines were extracted from the VALD database. For the Ni\,\textsc{i} 630.03 nm line we use $\log gf=-2.11$ from \citet{Johansson2003}.

For the NLTE calculations we employ a 23 level O atom (22 O\,\textsc{i} levels and the ground state of O\,\textsc{ii}), including in total 43 bound-bound and 22 bound-free transitions. This atom is based on the atom of \citet{Kiselman1993}, but has undergone updates \citep{Fabbian2009} as new data for electron-impact excitation has recently become available \citep{Barklem2007}.

\subsection{Image degradation and instrumental profiles\label{sec:psf}}

\begin{figure} 
  \centering
  \includegraphics[width=0.5\textwidth]{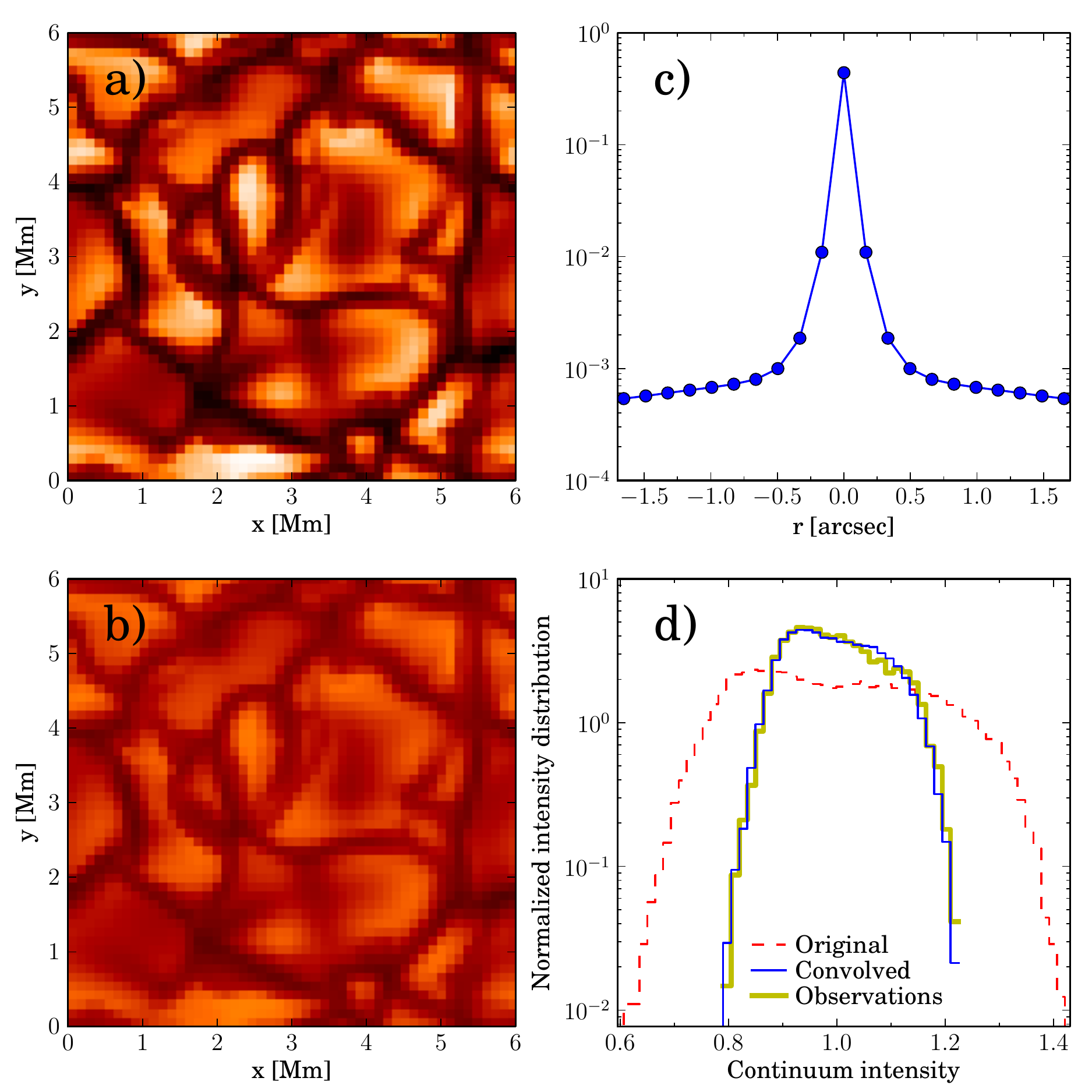}
  \caption{Convolution of the simulations with the telescope's and the Earth's atmospheric PSF at the solar disk-centre: a) Continuum intensity at 615 nm for the original simulation; b) Continuum intensity at 615 nm for the convolved simulation; c) The middle row of the 2D PSF used for the convolution (converted from simulation pixels to arcsec and in log scale); d) normalized continuum intensity distribution at 615 nm for the simulations and observations. The convolution parameters $a$ and $b$ (see text) are chosen so that the continuum intensity distribution and continuum contrast of the simulations match the observations.}
  \label{fig:convolution}
\end{figure}
To compare the simulations with Earth-based observations one needs to degrade them in a similar manner as the Earth's atmosphere and the telescope degrade the images from the Sun. Additionally, the spectra will also have to be convolved to account for the spectrograph's instrumental profile. We employ a three step convolution process that mimics the path of light from the solar surface to the spectrograph: \mbox{atmosphere $\to$ telescope $\to$ spectrograph}. The first step is the point-spread function (PSF) caused by the Earth's atmosphere. Following \citet{Nordlund1984} and \citet{ColladosVazquez1987} we describe the atmospheric PSF as a sum of two Lorentzians:
\begin{equation}
  \label{eq:lorentz}
p_{\mathrm{atm}}(r,a,b) = \frac{a}{r^2+a^2}+\frac{b}{r^2+b^2},
\end{equation}
where $r$ is angular distance and we define $a > b$ so that the first Lorentzian determines the overall contrast (`extended wings' of the PSF) and the second Lorentzian determines the size of the smallest detail observable (`narrow core' of the PSF). Tests using different functions for the PSF show that the effect on the results is negligible. 

Following the path of light, the next step is to mimic the finite resolution of the telescope. We employ an Airy disk function for a circular aperture telescope as the `diffraction' PSF:
\begin{eqnarray}
  \label{eq:airy}
x &=& \pi\frac{D}{\lambda}r \\
p_{\mathrm{diff}}(x) &=& \left[2\frac{J_1(x)}{x}\right]^2,
\end{eqnarray}
where $D$ is the telescope's diameter, $\lambda$ the wavelength of the observations, $r$ the angular distance and $J_1$ a Bessel function of the first kind.

The combined PSF is then the convolution of $p_{\mathrm{atm}}$ with $p_{\mathrm{diff}}$:
\begin{equation}
  \label{eq:psf}
p(r,a,b) = \left[p_{\mathrm{atm}} \ast p_{\mathrm{diff}}\right](r,a,b).
\end{equation}

The PSF is normalized, so that the convolution with the images conserves the intensity. The $a$ and $b$ parameters depend on the local conditions and are empirically obtained. Two properties of the observations are used to constrain them: the continuum intensity contrast and the continuum intensity distribution. These two properties are not completely independent of each other: a change of the distribution will change the contrast and vice-versa. Nevertheless it is possible, for a fixed contrast, to change the shape of the distribution using different pairs of $(a,b)$, and that is the approach taken. For the disk-centre data we adjust $a$ and $b$ in order to get the best agreement with the observed continuum intensity distribution and continuum contrast. Fig.~\ref{fig:convolution} shows an example of how 777 nm data is convolved so that it agrees with the observations. By fitting the observed intensity distributions, the adopted PSF formulation also corrects for the non-ideal components of the telescope and instrument PSFs. %

\begin{figure} 
  \centering
  \includegraphics[width=0.35\textwidth]{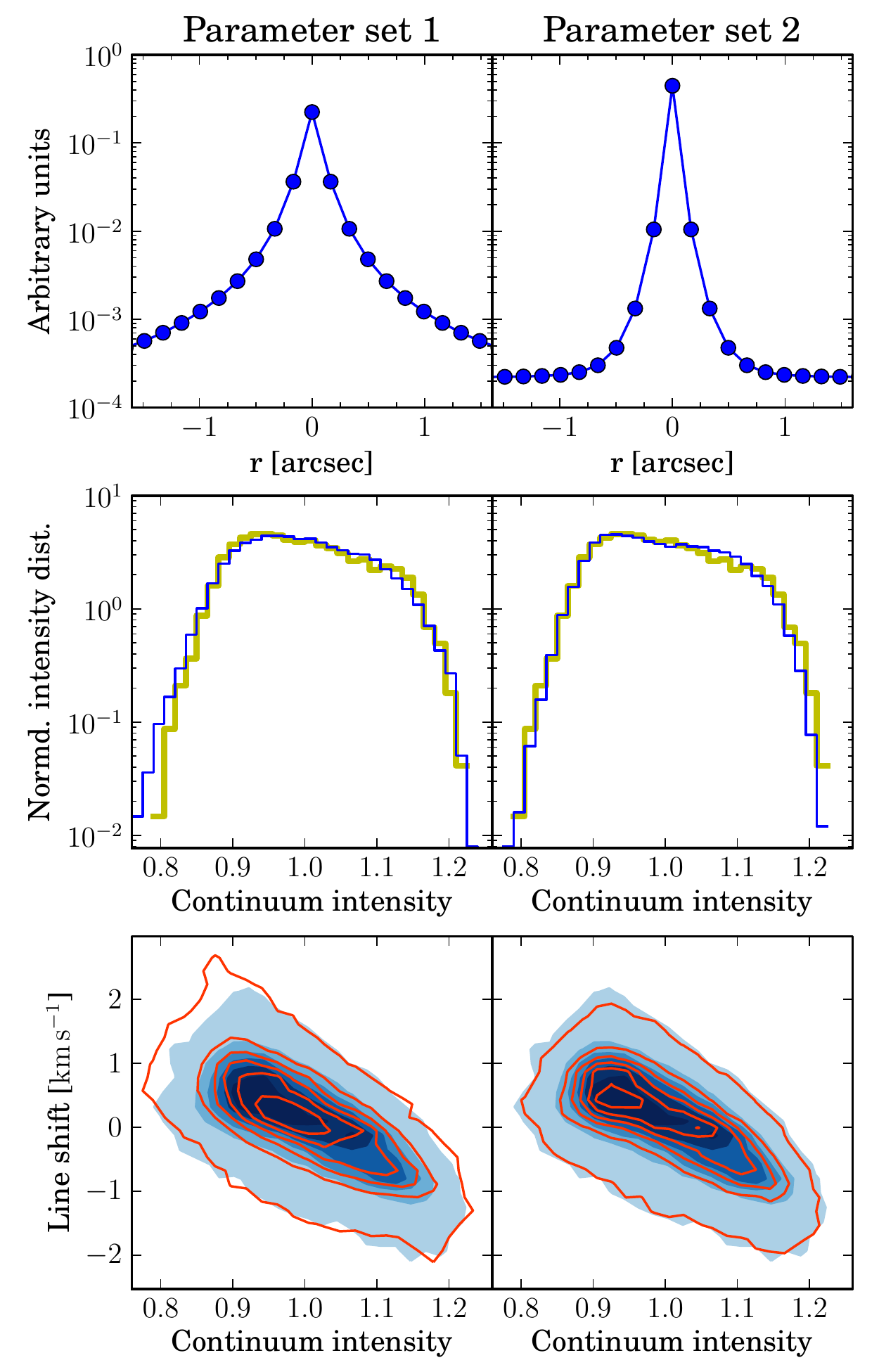}
  \caption{Effects of different PSF shapes at 615 nm, $\mu=1$, using two sets of $a$ and $b$ parameters (left and right columns). \emph{Top panels}: middle row of the 2D PSF. \emph{Middle panels}: continuum intensity distribution for the convolved simulations (thin line) and the observations (thick line). \emph{Bottom panels}: effects of the different convolutions in the line shifts of the strong Fe\,\textsc{i} 615.77 nm line. Contours represent the point density (see Sect.~\ref{sec:shifts}). Solid colour contours are for the convolution with the PSF adopted for our analysis (see Fig.~\ref{fig:convolution}), line contours for the convolution with the PSF of the corresponding parameter set.}
  \label{fig:psf_params}
\end{figure}

There is some degree of degeneracy in the intensity distribution for different values of $a$ and $b$. The adopted PSF is not necessarily the most physically correct. It is rather an empirical transformation that best reproduces the observations, given the coarse spatial resolution of the simulations and the time-average of different seeing conditions that the observed intensity distribution represents. Several tests were made to study the effect of different PSFs in the results. In Fig.~\ref{fig:psf_params} we plot some results for two sets of parameters giving different PSFs: one with stronger wings (left column), and the other with weaker wings (right column). The difference in the intensity distributions given by these two PSFs is very small, as is the effect on the results (in this case, plotted for the line shifts, but the results are similar for other parameters). We conclude that the adopted PSF describes the observations reasonably, and the error introduced in the results by a different shape of PSF is small. 

\citet{SteinNordlund1998} hint at a good agreement between the 3D model intensity distibutions and the observations. Recent work has indeed shown that the intensity distribution of these and other simulations are consistent with observations once careful PSF modelling is performed \citep{Wedemeyer2009,Danilovic2008}. \citet{Wedemeyer2009} also show that the observed centre-to-limb variation of the continuum contrast is well described by a 3D model of the Stein \& Nordlund family.. Nevertheless, and for a fair comparison, we chose not to assume too much about the predicted intensity distributions at $\mu\neq 1$. Unlike at disk-centre, we do not force the model intensity distribution to match the observations. The parameter $a$ is related to the PSF wings. Its sources are several: straylight in the telescope, atmospheric scattering, etc. It is reasonable to assume that it will not change significantly between observing sets. The parameter $b$, on the other hand, defines the seeing-induced resolution, which is expected to vary between sets. Hence for the $\mu\neq 1$ sets we use the same $a$ value as obtained for disk-centre, and adjust only $b$ to obtain the observed continuum contrast.

The simulation used has a horizontal resolution of $50\times 50$. For disk-centre this corresponds to a spatial scale of $6\times 6\: \mathrm{Mm}$. Because the pixel size of the box is kept constant, when one computes line profiles at $\mu\neq 1$ the spatial resolution will increase with decreasing $\mu$. Different $\varphi$ angles at off-centre positions will also cause different spatial resolutions along different axes of the simulation box. The two-dimensional PSF is adjusted so that the angular coordinate $r$ matches the spatial coordinate in pixels, meaning it has an ellipsoidal profile to account for $\mu$, rotated according to the $\varphi$ angle.

After convolving the synthetic spectrograms with the PSF, each individual spectrum is also convolved with the instrumental profile of the spectrograph. A Gaussian equivalent to a resolving power of 200\,000 was used. This value of the resolution was obtained by investigating the data and comparing with the FTS intensity atlas.

\section{Results and discussion\label{sec:results}}

\subsection{Line shifts\label{sec:shifts}}

\begin{figure*}[htb!]
\begin{center}
  \includegraphics[width=\textwidth]{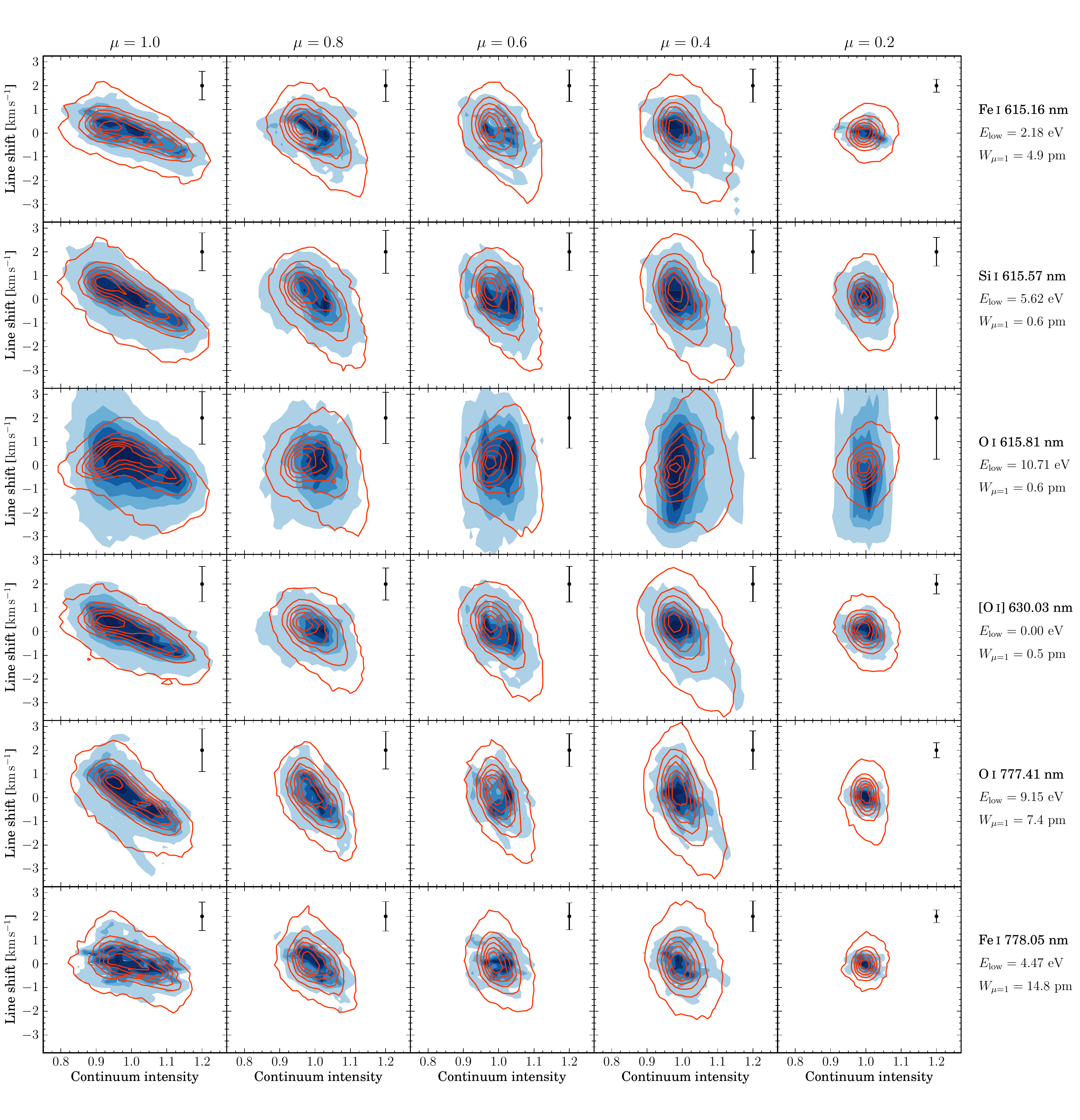}
\end{center}
\caption{Line shifts for selected lines and several $\mu$ values. Data shown as contour levels of the two dimensional histogram of the density of points. Observations are depicted as solid colour contours, while the 3D model as line contours. The continuum intensity is normalized to the mean intensity for each $\mu$ value. The lower continuum contrast towards the limb is evident, with less structure information at $\mu=0.2$. The energy of the lower level is shown, along with the disk-centre equivalent width. The chosen lines cover a representative range of lower level energies and equivalent widths. The results for O\,\textsc{i} 615.81 nm are uncertain due to the line being particularly weak, especially at intergranular regions and at the limb. Results for the other two O\,\textsc{i} 777 nm triplet lines are similar those shown for O\,\textsc{i} 7774.1 nm.}
  \label{fig:lineshifts}
\end{figure*}

The line shifts are a diagnostic of the velocity fields on the solar surface. Observations at disk-centre probe the vertical motions, while observations closer to the limb can also provide some information about horizontal motions. It should be noted, however, that because line formation takes place at a range of depths it is not possible to associate the observed line shifts with a specific geometric height. The lower contrast of the off-centre data (especially at the limb) means that a precise comparison of the line-of-sight velocities between the different positions on the solar disk is very difficult. 

Some results are plotted in Fig.~\ref{fig:lineshifts}. The different lines have different formation ranges, and the line shifts give us information about the velocity fields where these lines were formed. For the line shifts as well as most of the other line information, the disk-centre data provide the most detail in the structure as a function of continuum intensity. One can see very clearly the trend with the continuum intensity: the dark (intergranular) regions representing downflows and the bright (granules) regions representing upflows. The agreement with the 3D model is very good, confirming its realistic velocity fields. 

The results for other lines are very similar to Fig.~\ref{fig:lineshifts}, with generally an excellent agreement between the 3D model and the observations. For some very weak lines with close stronger lines (\emph{e.g.} O\,\textsc{i} 615.81 nm) our algorithm for determining the line centre can fail more often. This is due to the low S/N ratio and difficulty in finding a very weak line, and results in an increased scatter in the line shifts.

\subsection{Line strengths}

The distribution of the line equivalent widths over the solar granulation is one of the most interesting diagnostics. It provides a wealth of information on how lines are formed over a depth range. From an observational point of view it can be used to `fingerprint' particular types of species \citep{KiselmanAsplund2003}. But more interestingly, it can be used to probe the structure of 3D models and the microphysics of line formation. The early work of \citet{Kiselman1998} is an example of how similar observations can be used to probe for departures from LTE in the 671 nm Li\,\textsc{i} line. Indeed in Sect.~\ref{sec:nlte} we detail a similar analysis of the NLTE effects of the O\,\textsc{i} lines using the spatially resolved variation of equivalent widths.

\begin{figure*}
\begin{center}
  \includegraphics[width=\textwidth]{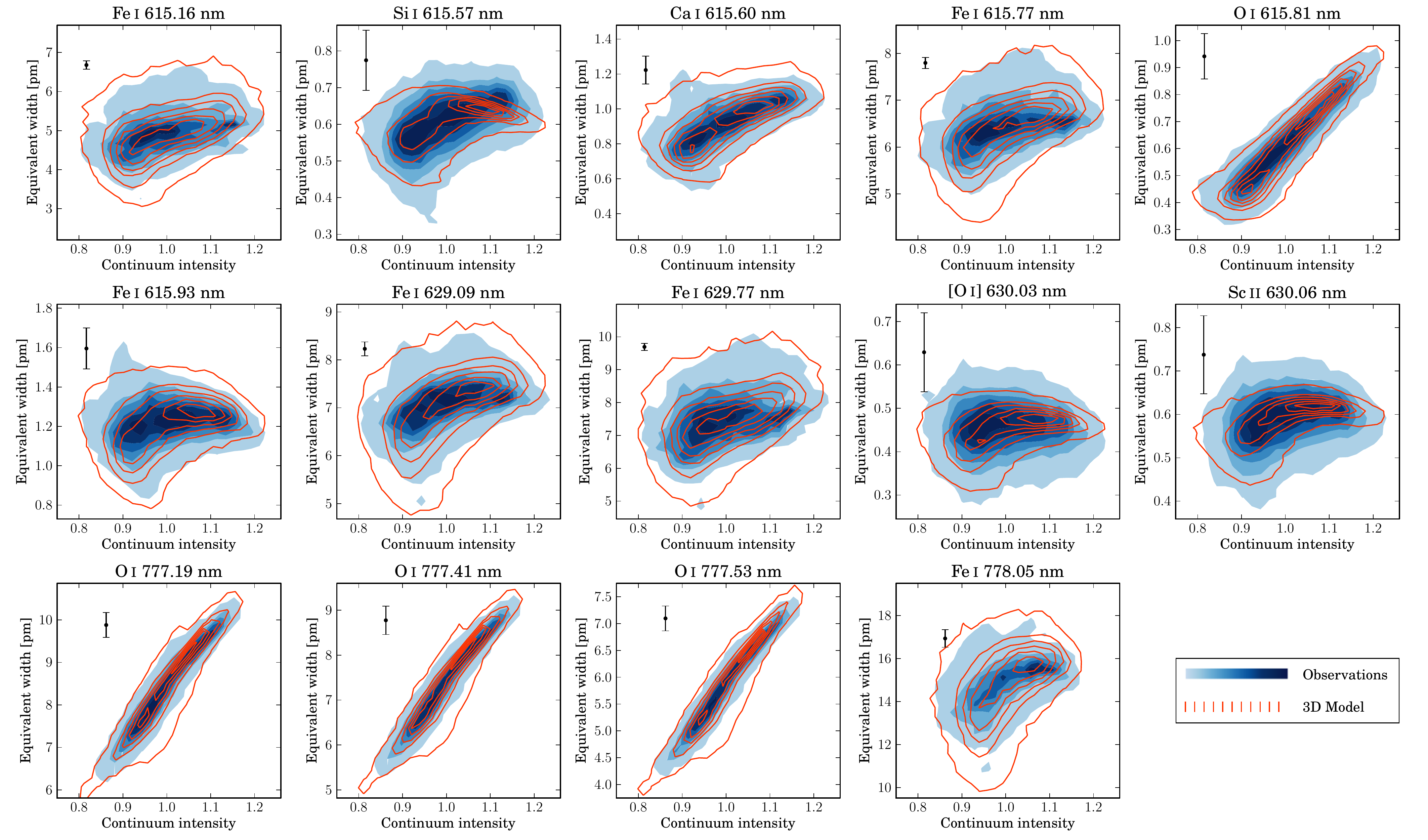}
\end{center}
\caption{Distribution of equivalent widths over the solar granulation at disk-centre. For each spatially-resolved spectrum the equivalent width has been computed and the resulting histogram of points is represented by the contours, as a function of the normalized local continuum intensity. The line profiles of Si\,\textsc{i} 615.57 nm, Ca\,\textsc{i} 615.60 nm, O\,\textsc{i} 615.82 nm and [O\,\textsc{i}] 630.03 nm include nearby blends (see Sect.~\ref{sec:atomd}). All line profiles were computed assuming LTE, except the O\,\textsc{i} 777 nm lines, showing the NLTE results for $S_\mathrm{H}=1$.}
  \label{fig:eqw_mu1}
\end{figure*}

In Fig.~\ref{fig:eqw_mu1} we show the distribution of equivalent widths for all the lines considered in the present study. The contours reflect the density of points in the plot. Statistically observations and 3D model are comparable, the contours comprising $\approx$50\,000 spatial points, exception made for the O\,\textsc{i} 777 nm lines, where the contours from the NLTE profiles comprise 25\,000 points (only 10 snapshots were used for NLTE calculations).

With a relatively low signal-to-noise ratio for each spatial point, it is not surprising that the observations of weak lines show a bigger scatter than the 3D model. Photon noise influences the equivalent widths directly and also the determination of the continuum level for each spatial point, effects that become more pronounced in weaker lines. %

Overall the agreement between the 3D model and the observations is very good. The Si\,\textsc{i} 615.57 nm line shows a slight mismatch, with the model predicting a weaker line at the bright end. This may be due to blend contamination. In the line profile we have included blends with a nearby strong Si\,\textsc{i} line and also the Ca\,\textsc{i} 615.60 nm (because their wings overlap), but the problem persists. In this region the VALD database lists only a very weak Fe\,\textsc{i} which is unlikely to alleviate the problem either. We note that the Ca\,\textsc{i} 615.60 nm would display a similar behaviour if we had not included the blend with the O\,\textsc{i} 615.6 nm line ($E_\mathrm{low}$ = \mbox{10.74 eV}). For the Si\,\textsc{i} it is possible that a similar high-excitation blend might be at play. 

For the other lines it is interesting to note the effect of the excitation energy: the O\,\textsc{i} 777 and 615.82 nm having a high excitation energy and a very pronounced equivalent width variation, while other lines have a more `C' shaped pattern, becoming increasingly flat as the excitation energy decreases (\emph{e.g.} [O\,\textsc{i}]). The variation of the equivalent width distribution with excitation energy is discussed in Sect.~\ref{sec:variations}. For the strong Fe\,\textsc{i} lines the 3D model seems to predict a slightly steeper variation of equivalent width with continuum intensity at the bright end. This is visible for the Fe\,\textsc{i} 615.77 nm, 629.09 nm, 629.77 nm and, to a lesser extent, 778.05 nm. The origin of this effect is unclear. 

The line profiles were adjusted in abundance so that they match the observed $W_{\mathrm{mspec}}$. This adjustment essentially moves the distribution vertically in Fig.~\ref{fig:eqw_mu1}. Again, it is very reassuring to see how well the 3D model and LTE line formation (except O\,\textsc{i} 777 nm) describe the observed distributions. 

\subsection{Line asymmetries}

\begin{figure}
\begin{center}
  \includegraphics[width=0.49\textwidth]{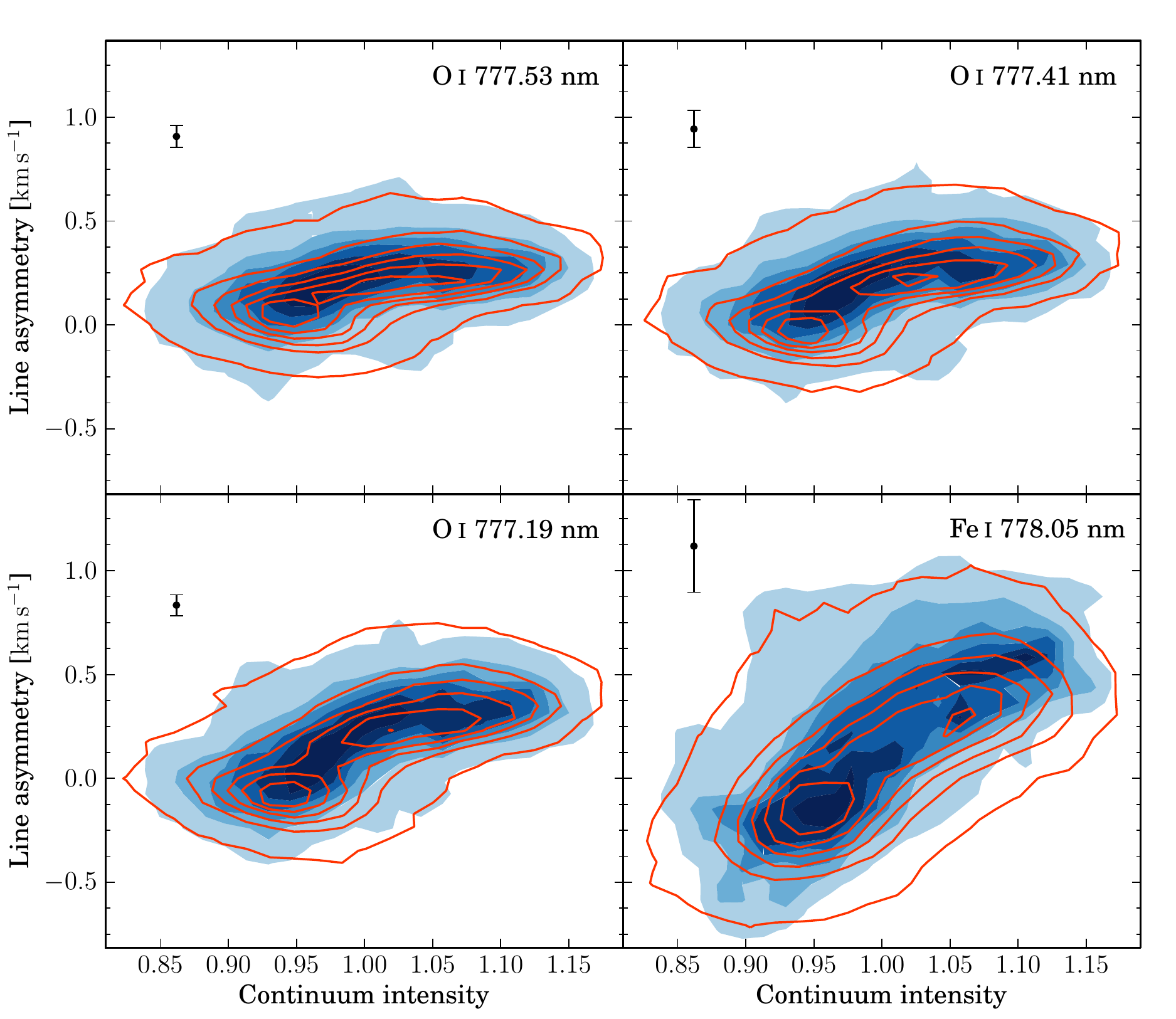}
\end{center}
\caption{Line asymmetries at disk-centre for four of the strongest lines studied: the O\,\textsc{i} 777 nm triplet and Fe\,\textsc{i} 778.05 nm. Comparing observations (solid colour contours) with 3D model (line contours). For O\,\textsc{i} 777 nm showing results from LTE profiles due to higher number of snapshots computed (the asymmetry results are not very sensitive to departures from LTE).}
  \label{fig:asym_spatial}
\end{figure}

The line-asymmetry measure is defined here as the wavelength difference (in velocity units) between the line core and the bisector at half maximum. It is a two-point sampling of the line bisector and, along with the line shift, another probe of the velocity fields. For a given point in the solar granulation, the line asymmetry reflects the velocity distribution over its formation depth and optical path. %

\begin{figure}
\begin{center}
  \includegraphics[width=0.49\textwidth]{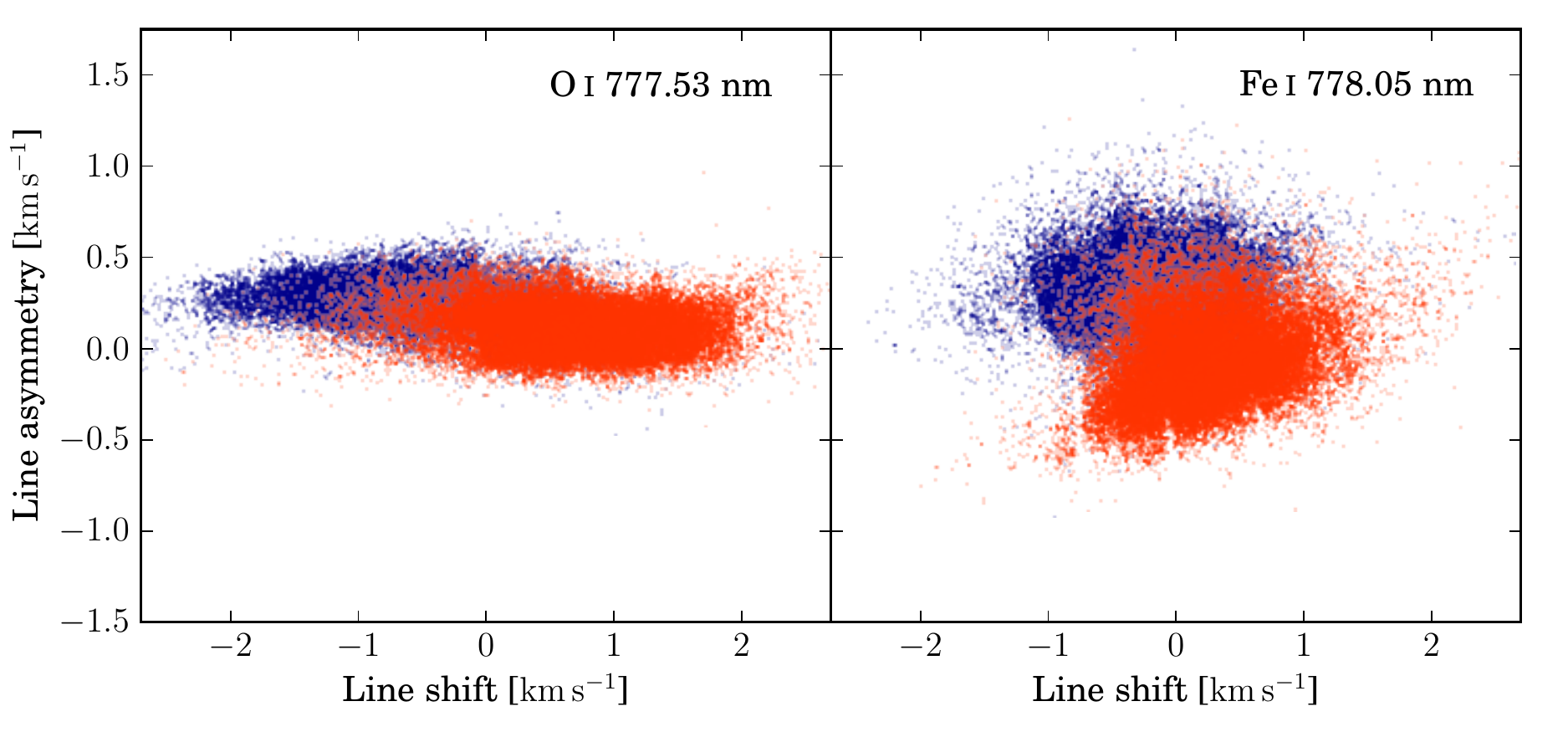}  
\end{center}
\caption{Line asymmetries vs. line shifts for the 3D model, LTE, disk-centre and convolved to match the observations. For clarity plotted with points instead of contours. Dark points represent $I_{\mathrm{c}} > \langle I_{\mathrm{c}}\rangle$, light points $I_{\mathrm{c}} < \langle I_{\mathrm{c}}\rangle$.}
  \label{fig:asym_lc}
\end{figure}

In Fig.~\ref{fig:asym_spatial} we show the variation of line asymmetry at disk-centre for four medium to strong lines. For weak lines the asymmetry distribution is essentially flat, meaning that the velocity differences along the formation region are small. The O\,\textsc{i} 777 nm lines provide a great probe for the line asymmetries, because they have essentially the same line parameters except the $\log gf$ value. In Fig.~\ref{fig:asym_spatial} we can see how the asymmetry distribution changes from the the weakest line (777.53 nm) to the strongest (777.19 nm). For the weakest line there is almost no variation with continuum intensity, and an overall positive asymmetry. This means that the `bulk' of the line profile is blueshifted in regards to the line core, a result of the vertical velocities in its formation region being mostly positive (outgoing material). As the lines get stronger their formation range will extend outwards. The vertical velocities in higher layers of the photosphere tend to get progressively smaller (with an increase of the horizontal motions as the gas outflows cool down). Formed at higher depths, the stronger lines will cover a larger velocity gradient in the photosphere, and a trend of asymmetry with continuum intensity becomes noticeable. The dark intergranular regions will have a negative asymmetry (the line profile is redshifted in relation to the core) and the brighter regions will have a positive asymmetry (the line profile is blueshifted in relation to the core). These effects are a result of the downflows and upflows of gas at these regions. They become even more apparent in the much stronger Fe\,\textsc{i} 778.05 nm line, covering a significant velocity gradient and displaying the most extreme values of asymmetry from all lines.

For these four lines as well as most of the other lines (not shown) the 3D model predictions regarding line asymmetry are very good. This is not surprising, given the excellent results reported earlier about the line bisectors \citep{Asplund2000}. Our results for line asymmetries are consistent with the findings of \citet{Kiselman1994}.

In Fig.~\ref{fig:asym_lc} a different perspective is shown. Here the asymmetries of the weakest of the O\,\textsc{i} 777 nm lines and the strong Fe\,\textsc{i} line are plotted against the line shifts (measured from the line core in relation to the mean spectrum). The points are colour-coded between dark and bright continuum intensity regions. For the weaker line the average line shift for the bright granules is negative, and positive for intergranular regions. For the strong line granules and intergranular lane regions occupy roughly the same space in line shift, and the overall dispersion in line shifts is smaller. One explanation for this could be that the core of this line is formed above the convective overshoot region. But while the core is formed higher, the asymmetry of the line profiles reflects the motions below, and it becomes very clear that the negative asymmetry is associated with dark regions and positive asymmetry with bright regions. This distinction is hardly visible in the weaker line, formed deeper in the photosphere.

\subsection{Line FWHM\label{sec:fwhm}}

\begin{figure}
\begin{center}
  \includegraphics[width=0.49\textwidth]{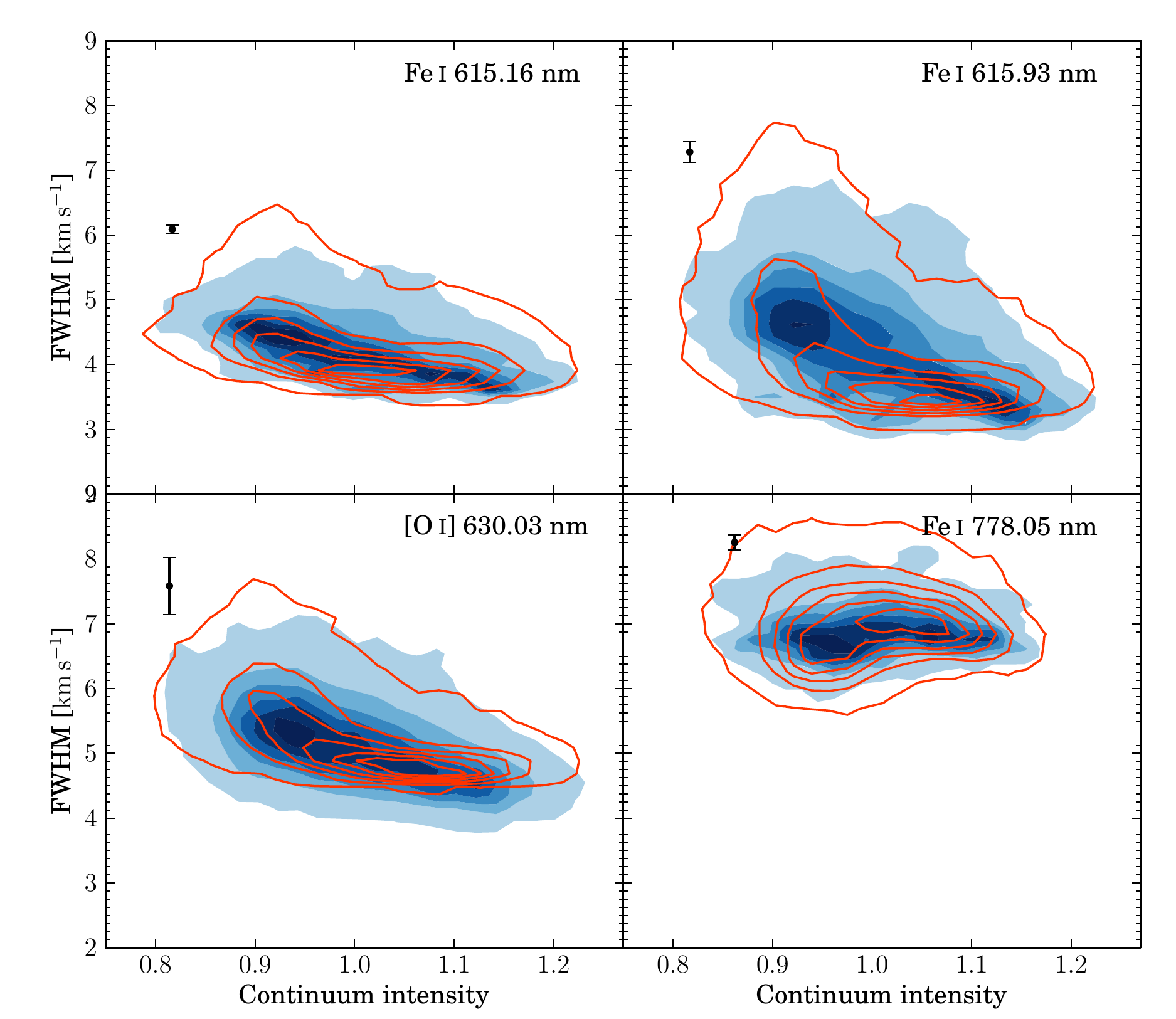}
\end{center}
\caption{Line full widths at half maximum (FWHM) at disk-centre for four lines of different strengths and excitation energies. Comparing observations (solid colour contours) with 3D model (line contours).}
  \label{fig:fwhm_spatial}
\end{figure}

The distribution of the FWHM with continuum intensity for disk-centre is shown in Fig.~\ref{fig:fwhm_spatial}, for four lines. For most of the lines the FWHM decreases with increasing continuum intensity, with a larger scatter at low continuum intensity. This decreasing behaviour of FWHM with continuum intensity has been observed by others \citep[\emph{e.g.}][]{Hanslmeier1993,Kiselman1994,Puschmann2003} and is caused by increased line broadening due to turbulence in the intergranular regions. For example, tests with the [O\,\textsc{i}] 630~nm line show that if the velocity fields are switched off when computing the theoretical profiles, then the predicted FWHM distribution is flat. Very strong lines like Fe\,\textsc{i} 778.05~nm are formed higher in the atmosphere and above the convective overshoot region, leading to a smaller turbulent broadening and a flatter FWHM distribution, as seen in Fig.~\ref{fig:fwhm_spatial}. 

The O\,\textsc{i} 777~nm lines are a special case in the FWHM distribution. Unlike all the other lines considered here, their net variation of FWHM is an increase with the continuum intensity. It can be seen in the bottom panel of Fig.~\ref{fig:sh_spatial} for O\,\textsc{i} 777.41~nm. This effect is related to thermal broadening and results from a peculiar combination of strong lines with a high excitation potential and the relatively light atom of oxygen. With \mbox{$E_{\mathrm{low}}=9.15$ eV} the O\,\textsc{i} 777~nm lines are formed very deep where the temperature is higher. A high temperature together with the relatively small atomic weight of oxygen causes the large thermal broadening seen in these lines, which dominates over turbulent broadening. With a steeper temperature gradient in the granules the thermal broadening is larger in the bright regions, driving the FWHM increase with continuum intensity. Tests with fake lines show that the negative slope of the FWHM with continuum intensity can be recovered if the atomic weight is assumed to be much higher than that of oxygen (thereby minimizing thermal broadening), confirming that the FWHM trend is dominated by thermal broadening. 

Another candidate for this kind of behaviour is the O\,\textsc{i} 615.81~nm (\mbox{$E_{\mathrm{low}}=10.74$ eV}). However, because this line is much weaker, in absolute terms its thermal broadening will be much smaller, and turbulent broadening dominates. The model predictions (not shown) confirm a decrease of FWHM with the continuum intensity. The inclusion of the molecular blends around this line (see discussion in \citetalias{Pereira2009a}) will reinforce even more the negative variation of FWHM with continuum intensity. In the observations the uncertainty in its FWHM is very large because the line is very weak, preventing a confrontation with the model predictions.

Overall the 3D model reproduces the observed distributions of FWHMs very well, for both strong and weak lines and including the special case of the O\,\textsc{i} 777~nm lines. %

\subsection{Departures from LTE in oxygen lines\label{sec:nlte}}

Of the O\,\textsc{i} lines the 777~nm triplet lines are know to suffer significant departures from LTE. This was first shown by \citet{Altrock1968} and later explained by \citet{Eriksson1979}. Subsequent studies \citep[\emph{e.g.}][]{Kiselman1991} found also NLTE effects for the O\,\textsc{i} 615.81~nm line, albeit on a much smaller scale -- usually $\lesssim 0.03$ dex (\citealt{Asplund2004,Caffau2008};\mbox{\citetalias{Pereira2009a}}).

A proper NLTE treatment of these lines poses several challenges. As noted in previous studies, one of the main problems is to account for the effect of collisions with neutral hydrogen. No laboratory data or quantum mechanical calculations are available to date. The best estimate to account for the efficiency of hydrogen collisions has been to use the recipe of \citet{Steenbock1984}, a generalization of the classical Drawin formula \citep{Drawin1968}. This recipe is often scaled by an empirical factor $S_\mathrm{H}$. In the work of \citet{CAP2004} and \mbox{\citetalias{Pereira2009a}} the centre-to-limb variations of the O\,\textsc{i} 777 nm lines have been used to derive the best-fitting $S_\mathrm{H}$ value. 

We propose to use the line variations across the solar granulation at disk-centre to test the LTE and NLTE line formation and derive the best fitting empirical $S_\mathrm{H}$ value. The inherent scatter of the spatially-resolved line quantities makes this method less sensitive or constraining than the centre-to-limb variations. Nevertheless, it provides an independent test of the possible $S_\mathrm{H}$ values. 

We have used the same LTE and NLTE calculations as in \citetalias{Pereira2009a}, detailed in Sect.~\ref{sec:line_formation}. Due to computational constraints the full 3D NLTE line formation was only performed for 10 snapshots of the 3D simulation. The NLTE effects in O\,\textsc{i} 615.81 nm are so small that the distributions of line quantities in NLTE are essentially indistinguishable from LTE. Hence we limit our NLTE analysis to the 777 nm triplet only. The line profiles for different values of $S_\mathrm{H}$ have all been interpolated in abundance so that the equivalent width of the spatially and temporally averaged spectrum matches the observed.

\begin{figure}
\begin{center}
  \includegraphics[width=0.49\textwidth]{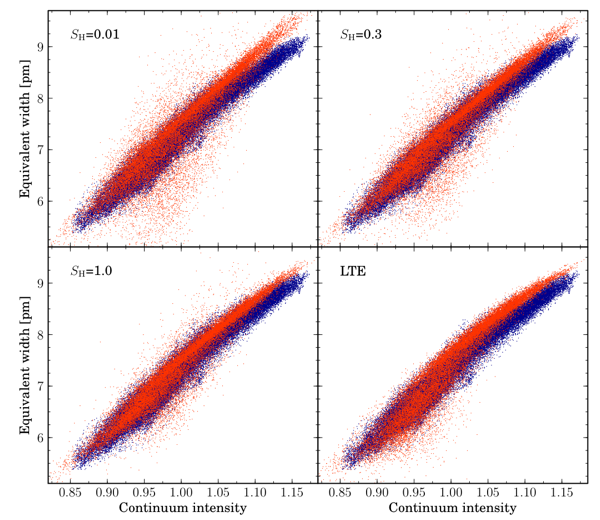}
  \includegraphics[width=0.49\textwidth]{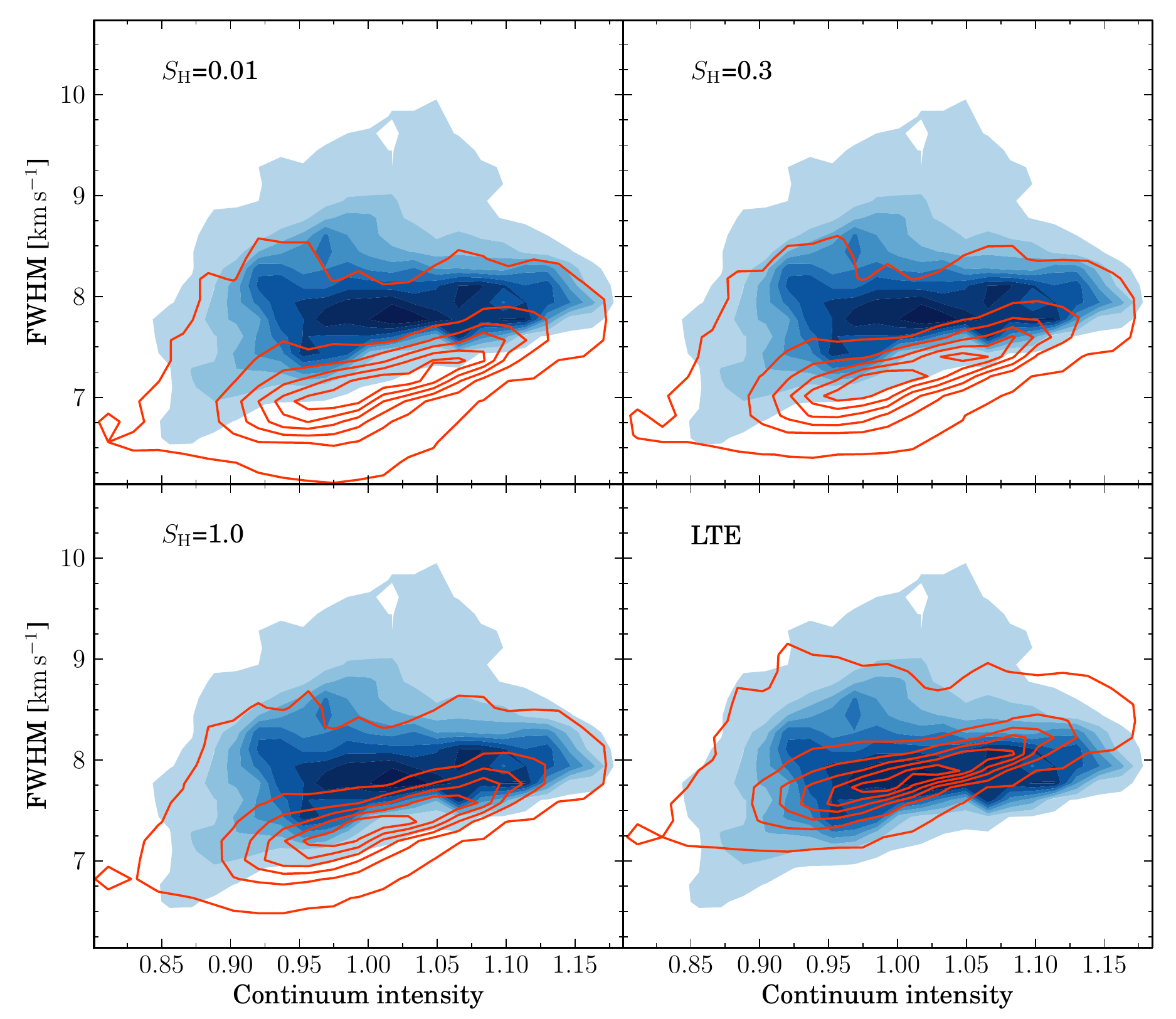}
\end{center}
\caption{\emph{Top:} equivalent widths of the O\,\textsc{i} 777.41 nm line as a function of continuum intensity. For clarity the values are shown as points, instead of contours. \emph{Bottom:} FWHM for the same line as a function of continuum intensity. Comparing the disk-centre observations (dark points in equivalent width; solid color contours in FWHM) with the 3D model (light points in equivalent width, line contours in FWHM), for LTE and NLTE, for different values of $S_{\mathrm{H}}$. The oxygen abundance has been adjusted in each case to fit the observed mean disk-centre equivalent width.}
  \label{fig:sh_spatial}
\end{figure}

In Fig.~\ref{fig:sh_spatial} we show how the line equivalent widths and the FWHM vary across the solar granulation at disk-centre, for the observations and the 3D model with different $S_\mathrm{H}$ values for hydrogen collisions. The results shown are for the O\,\textsc{i} 777.41 nm line, but they are very similar for the other two lines. 

The equivalent width plot is perhaps the most interesting. The model results can be divided in two regions: granules and intergranular regions, with a smaller point density in the middle, where $I_{\rm{c}}=\langle I_{\rm{c}}\rangle$. The hydrogen collisions have different effects on each region. An increase in $S_\mathrm{H}$ makes the equivalent width slope less steep in the intergranular regions, but steeper in the granular regions. In the LTE case it can be seen that in the intergranular regions the equivalent width slope is steeper than in the observations. In such regions the agreement with the observations improves for the NLTE models. On the other hand, as $S_\mathrm{H}$ decreases the equivalent width slope in the bright regions gets steeper, departing from the observed values. No single value of $S_\mathrm{H}$ can be in perfect agreement with the observations in both high and low intensity regions. 

Another aspect worth noting is the scatter, especially in the dark regions. For $S_\mathrm{H}=0.01$ one can see that the 3D model displays a significant amount of scatter at low continuum intensity. This does not seem to be consistent with the observations. As $S_\mathrm{H}$ increases the amount of scatter decreases. This increase of scatter in NLTE is the opposite behaviour of the Li\,\textsc{i} 671 nm line: for this line there is a much bigger scatter for the LTE profiles \citep{Kiselman1998,Asplund2005}. Because the temperature dependence of the O\,\textsc{i} 777 nm lines is the opposite of Li\,\textsc{i} 671 nm, the opposite scatter results are perhaps not surprising. The increased scatter for low $S_\mathrm{H}$ is likely a reflection of the high temperature sensitivity of the lines. The NLTE effects make the lines stronger, and the lower $S_\mathrm{H}$, the stronger the lines. This strenghtening will increase the temperature sensitivity of the lines, which may explain the larger scatter for lower $S_\mathrm{H}$. If one makes the reasonable assumption that the 3D model realistically describes dark and bright granulation regions, then the scatter in equivalent width is enough to rule out the $S_\mathrm{H} \lesssim 0.3$ profiles, when comparing with the observations. Overall, accounting for the scatter and equivalent width slopes in dark and bright regions, $S_\mathrm{H}=1$ seems to be the value that best reproduces the observations. However, it should be noted that for $1 \leq S_\mathrm{H} \leq 3$ there is not much difference in the results.

In the lower panel of Fig.~\ref{fig:sh_spatial} we show a similar figure but for the FWHM. The area occupied by the observations's contours is larger probably due to a better statistical sample. Disregarding the line strength, the lower the $S_\mathrm{H}$, the narrower and deeper the NLTE line profile shapes will be when compared with LTE. Thus a decreasing $S_\mathrm{H}$ will drive down the FWHM. This is visible in Fig.~\ref{fig:sh_spatial}. The LTE FWHM's provide the best match for the observations. The NLTE profiles, even for $S_\mathrm{H}=1$ are narrower. This effect is also seen in the disk-centre profile fits of \citetalias{Pereira2009a}, and its cause is unclear. As in the equivalent width figure, lower $S_\mathrm{H}$ show a larger scatter in FWHM. However in this case the scatter cannot be used to exclude a particular $S_\mathrm{H}$ because the FWHM is always lower than the observations.

\subsection{Summary quantities of line variations\label{sec:variations}}

\subsubsection{Definitions}

In this section we look at the variation of the line quantities at high spatial resolution and at several $\mu$ values. It is not easy to quantify with a few numbers the quantity distributions like in Figs.~\ref{fig:lineshifts} and~\ref{fig:eqw_mu1}, but it must be done for achieving insight. There are many possible correlations between the different line quantities can be established. We follow the approach of \citet{Kiselman1994} and define the two quantities:
\begin{eqnarray}
  \alpha &=& \frac{\mathrm{d}I_{\mathrm{line\:\: core}}}{\mathrm{d}I_\mathrm{c}},\\
  \gamma &=& \frac{\langle I_\mathrm{c}\rangle}{\langle W\rangle} \cdot \frac{\mathrm{d}W}{\mathrm{d}I_\mathrm{c}},%
\end{eqnarray}
where $I_{\mathrm{line\:\: core}}$ is the absolute intensity of the line core, $W$ the equivalent width and $I_\mathrm{c}$ the continuum intensity. Both parameters are obtained by least-squares polynomial fits to the data. In the case of $\alpha$ a line is fitted, while for $\gamma$ a parabola is fitted. Exceptions to this are the O\,\textsc{i} 777 nm lines at $\mu=1$ and $\mu=0.8$. Because their equivalent width variations at these positions are not easily described by a parabola, 5th order polynomials are fitted instead. 

The $\alpha$ parameter is a measure of the variation of the absolute line core intensity with the local continuum intensity. If $\alpha=1$ then the line cores follow the continuum variations (expected for weak lines). Lower values of $\alpha$ are expected as the lines get stronger. This parameter is plotted vs. the equivalent width of each line (measured from the spatial mean spectrum at each $\mu$). 

The $\gamma$ parameter describes the variation of equivalent width with the local continuum intensity. It is divided by the mean equivalent width to put all lines in the same scale. To better describe the equivalent width distribution, the derivative $\mathrm{d}W/\mathrm{d}I_\mathrm{c}$ is evaluated at two points: in the dark (intergranular) regions and in the bright (granular) regions. These points are defined as the mean continuum intensity of all points where $I < \langle I_\mathrm{c}\rangle$ and $I > \langle I_\mathrm{c}\rangle$, respectively. Hence two values of $\gamma$ are defined: $\gamma_{\rm{dark}}$ and  $\gamma_{\rm{bright}}$.

\begin{figure}
\begin{center}
  \includegraphics[width=0.45\textwidth]{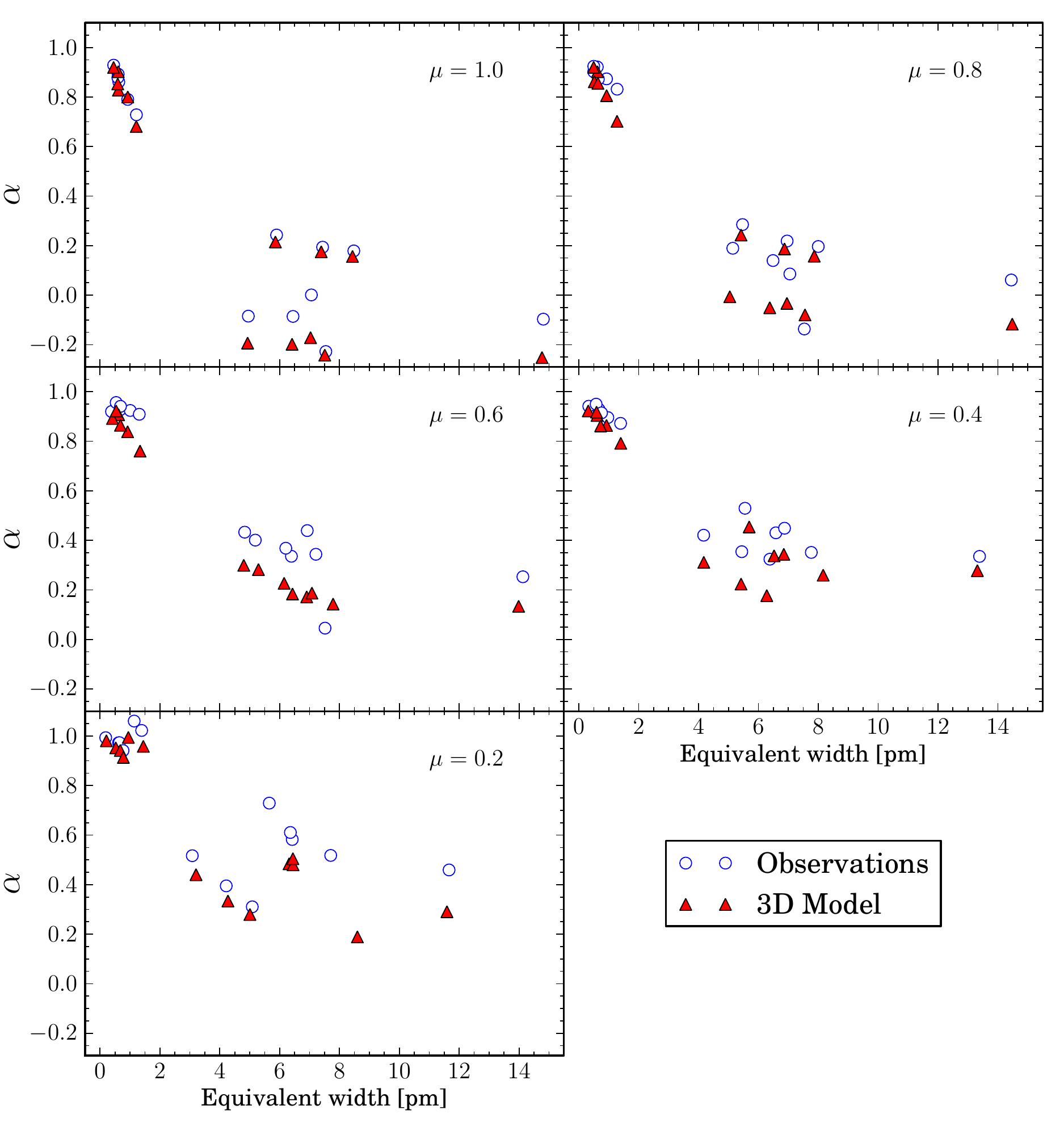}
\end{center}
\caption{Variation of the $\alpha$ parameter (describing variation of line core intensity, see text) with the mean equivalent width at each $\mu$, for all lines. The elemental abundances are adjusted so the model equivalent widths at $\mu=1$ match the observed. For \mbox{$\mu\neq 1$} the disk-centre abundances are used, meaning that the model equivalent widths depart from the observed in some cases.}
  \label{fig:alpha}
\end{figure}

\begin{figure}
\begin{center}
  \includegraphics[width=0.45\textwidth]{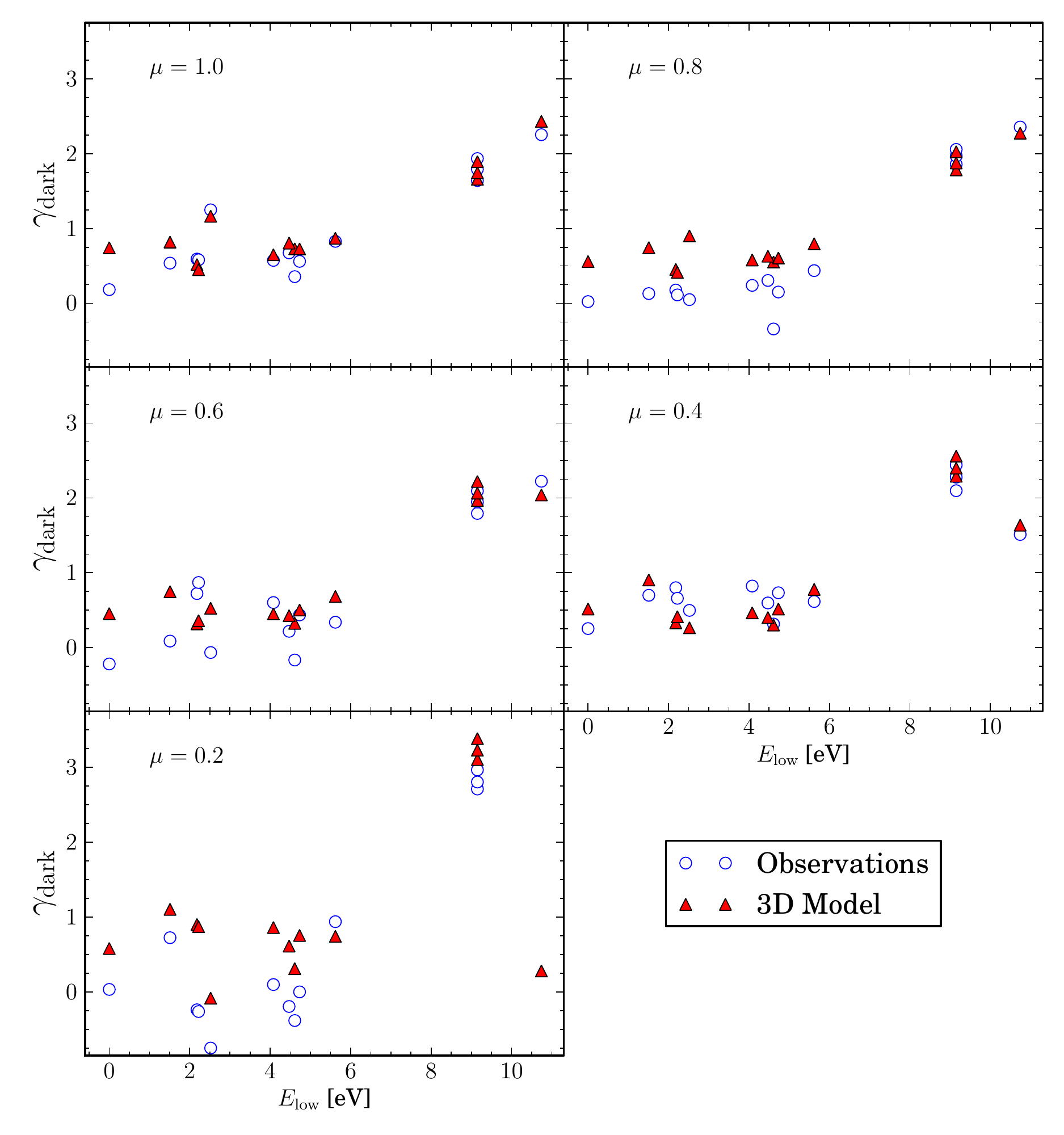}
  \includegraphics[width=0.45\textwidth]{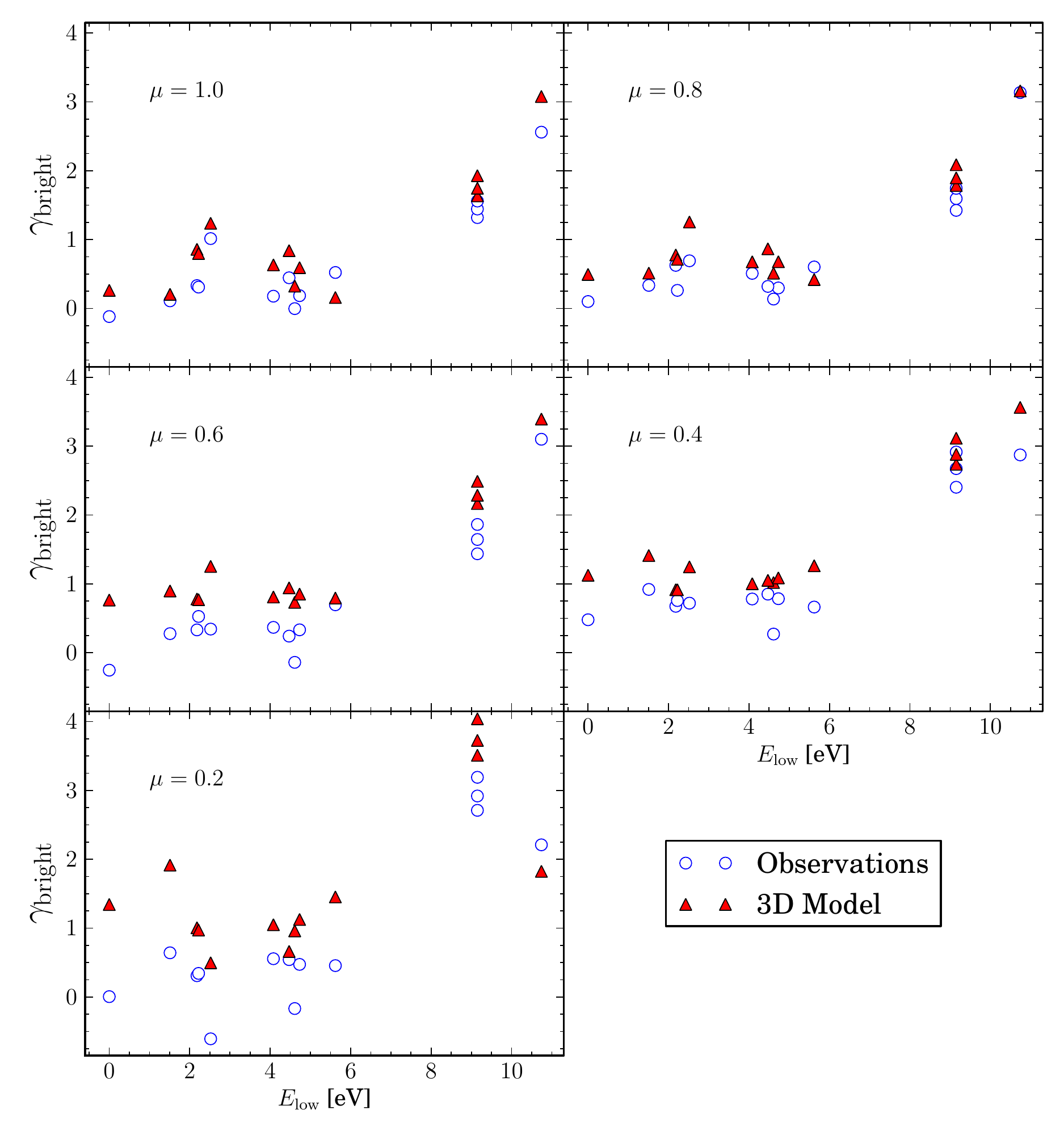}
\end{center}
\caption{Variation of the $\gamma$ parameter (describing the variation of equivalent width, see text) with the excitation potential for all lines, at the different $\mu$. The parameter is evaluated for dark (\emph{top panel}) and bright regions (\emph{bottom panel}). }
  \label{fig:gamma}
\end{figure}

\subsubsection{Results}

The results for $\alpha$ are plotted in Fig.~\ref{fig:alpha} and for $\gamma$ in Fig.~\ref{fig:gamma}. It should be noted that these quantities are subject to several uncertainties (particularly $\gamma$). The most important source of uncertainty is the scatter in the quantities and the approximation of using a line or parabola to fit these quantities. For the 3D models the particulars of the image degradation (see Sect.~\ref{sec:psf}) also influence these quantities, albeit to a lesser extent. The elemental abundances have been adjusted so that the model $W_{\mu=1}$ match the observations. The $\mu=1$ abundances are used for all the other $\mu$ positions. For the O\,\textsc{i} 777 nm lines the NLTE profiles with $S_{\rm{H}}=1$ were used.

The $\alpha$ results at $\mu=1$ are consistent with the findings of \citet{Kiselman1994}. The lack of lines with $W$ in the 1-5 pm region prevents a clear picture in the variation of $\alpha$ with equivalent width, but from the weaker lines a downward linear trend is visible. This goes on to a plateau for the stronger lines, where $\alpha$ is close to zero or negative. A negative correlation of the line core intensity with $I_{\rm{c}}$ for strong lines has been documented in the literature \citep[\emph{e.g.}][]{Balthasar1990,Hanslmeier1990,Kucera1995,Puschmann2003}. It is believed to be associated with convective overshooting: in higher layers the gas is cooler above granules than above the integranular lanes. This translates into an inverse granulation pattern, and causes the negative correlation between core intensity and $I_{\rm{c}}$. The model results indicate a slightly lower $\alpha$ for some of the stronger lines, which may indicate an higher convective overshooting than observed. For $\mu \neq 1$, $\alpha$ of the strong lines increases. Tests with the 3D model at $\mu=1$ and varying degrees of atmospheric degradation show that this effect (and also a less steep dependence of $\alpha$ with $W$) is mostly due to a reduction in continuum contrast. Overall the model agrees reasonably well with the observations. The stronger lines are more problematic, especially at the limb where the predicted equivalent widths differ from the observations.

The parameter $\gamma$ measures how the line strength varies with the continuum intensity. At $\mu=1$ we find both $\gamma_{\rm{dark}}$ and $\gamma_{\rm{bright}}$ to have a monotonic increase with the excitation potential of the lines. This is in agreement with \citet{Kiselman1994}, although he finds a more clear relation by employing a filter to remove the effects of solar oscillations. As discussed by \citet{Kiselman1994} this trend is an indication of the higher temperature sensitivity of lines with higher excitation potentials.  A positive $\gamma$ at $\mu=1$ is in agreement with the positive correlation between $W$ and $I_{\rm{c}}$ found in other studies \citep{Hanslmeier1990,Puschmann2003}. %

The variation of $\gamma$ between dark and bright regions is small but measurable. For most lines $\gamma_{\rm{dark}} > \gamma_{\rm{bright}}$, indicating a steeper variation of $W$ with  $I_{\rm{c}}$. The decrease in intensity contrast towards the limb (\emph{e.g.}, Fig.~\ref{fig:lineshifts}) makes the determination of $\gamma$ more problematic -- the uncertainty from fitting polynomials increases. Yet the agreement with the model is good for most lines and $\mu$ positions. It is interesting to note the agreement between model and observations is better for $\gamma_{\rm{dark}}$ than $\gamma_{\rm{bright}}$. The model predicts a steeper variation of $W$ with $I_{\rm{c}}$ when $I_{\rm{c}} < \langle I_{\rm{c}}\rangle$. For the O\,\textsc{i} 777 nm lines this is visible in Fig.~\ref{fig:sh_spatial}, where for LTE and different NLTE recipes, the model variation of $W$ in bright regions is always steeper than the observations. A higher $\gamma_{\rm{bright}}$ in the models when compared with the observations is more apparent for $\mu \la 0.6$.

\section{Conclusions\label{sec:conc}}

We have obtained high-spatial-resolution observations of lines of neutral oxygen and other species in quiet solar granulation at different positions on the solar disk. These have been used to test the 3D photosphere models and line formation. At the solar disk-centre the predicted distributions of line strengths, shifts and shapes are for most lines in excellent agreement with the observed. Both for the oxygen lines and most of the other lines. At different viewing angles a quantitative study of the line properties is more difficult because of a smaller contrast between granules and intergranular regions. Nevertheless, fits to the variation of line strengths and other quantities over the granulation show an encouraging good agreement with the observations.

For the O\,\textsc{i} 777 nm lines we employ the high-spatial-resolution observations at disk-centre to constrain the NLTE physics. Although not as robust as the centre-to-limb variation (\citealt{CAP2004};\citetalias{Pereira2009a}), this test provides another approach to empirically constrain the efficiency of collisions with neutral hydrogen, $S_{\rm{H}}$. We find that $S_{\rm{H}}=1$ describes most observables well, in agreement with the results of \citetalias{Pereira2009a}. Also in agreement with \citetalias{Pereira2009a} we find that NLTE profiles for $\mu=1$ are narrower than the observations, which agree better with the higher $S_{\rm{H}}$ profiles. Furthermore, it is difficult for any $S_{\rm{H}}$ to describe the variation in equivalent width with continuum intensity in the intergranular regions and granules at the same time. These discrepancies may be linked to the crude approximation of using scaled classical formul\ae{} to estimate the collisions with hydrogen. The discrepancies may indicate a more complex dependence of these collisions with temperature and/or pressure. Moreover, $S_{\rm{H}}$ will also likely vary between transitions.

Regarding the weak O\,\textsc{i} 615.81 nm line, its NLTE effects are very small and not noticeable in the variation of the line properties over the granulation. 

For the spectral lines investigated here, the agreement between the 3D model and the observations for the line velocities, shapes and strengths is excellent, both for disk-centre and different viewing angles. It places the 3D model on a solid footing regarding oxygen and provides compelling evidence on its suitability for abundance analysis.

\begin{acknowledgements}
We would like to thank C. Allende Prieto for providing us with his straylight code and P. S\"utterlin for his kind assistance with the Dutch Open Telescope. TMDP acknowledges financial support from Funda\c c\~ao para a Ci\^encia e Tecnologia (reference number SFRH/BD/21888/2005) and from the USO-SP International Graduate School for Solar Physics under a Marie Curie Early Stage Training Fellowship (project MEST-CT-2005-020395) from the European Union. This research has been partly funded by a grant from the Australian Research Council (DP0558836).
The Swedish 1-m Solar Telescope is operated on the island of La Palma by the Institute for Solar Physics of the Royal Swedish Academy of Sciences in the Spanish Observatorio del Roque de los Muchachos of the Instituto de Astrofísica de Canarias.
\end{acknowledgements}

\bibliography{3D}
\bibliographystyle{aa}

\end{document}